\documentclass[12pt, draftclsnofoot, onecolumn]{IEEEtran}
\usepackage{}
	\usepackage{pifont,bm,multicol,amsfonts,amsmath,color,amssymb,graphicx, epsfig,cite,psfrag,subfigure,algorithm,enumerate,stfloats,algorithm,algorithmic,epstopdf,balance,stfloats,mathrsfs}

    \usepackage{etoolbox}
    \usepackage{diagbox}
    \usepackage{graphicx}
    \makeatletter
    \patchcmd{\@makecaption}
     {\scshape}
     {}
     {}
    \makeatletter

\begin{document}

\title{Two-Bit RIS-Aided Communications at 3.5GHz: Some Insights from the Measurement Results Under Multiple Practical Scenes}

\author{Shun Zhang, \emph{Senior Member, IEEE}, Haoran Sun, Runze Yu, Hongshenyuan Cui, Jian Ren, \emph{Member, IEEE}, Feifei Gao, \emph{Fellow, IEEE}, Shi Jin, \emph{Senior Member, IEEE}, Hongxiang Xie and Hao Wang

\thanks{


Shun Zhang, Haoran Sun, Runze Yu and Hongshenyuan Cui are with the State Key Laboratory of Integrated Services Networks, Xidian University, Xi'an 710071, China (e-mail: zhangshunsdu@xidian.edu.cn, sdytshr@163.com, runzeyu@stu.xidian.edu.cn, hsycui@stu.xidian.edu.cn).

Jian Ren is with the National Key Laboratory of Antennas and Microwave Technology, Xidian University, Xi'an 710071, China (e-mail:
renjianroy@gmail.com).

Feifei Gao is with the Department of Department of Automation, Tsinghua University, Tsinghua University, Beijing, China (e-mail:
feifeigao@ieee.org).

Shi Jin is with the National Mobile Communications Research Laboratory, Southeast University, Nanjing 210096, China (e-mail: jinshi@seu.edu.cn).

Hongxiang Xie and Hao Wang are with the Huawei Technologies Co. LTD, Beijing 100095, China (e-mail: xiehongxiang@hisilicon.com, hunter.wanghao@hisilicon.com).

}
}

\maketitle

\vspace{-15mm}

\begin{abstract}
In this paper, we propose a two-bit reconfigurable intelligent surface (RIS)-aided communication system, which mainly consists of a two-bit RIS, a transmitter and a receiver.
A corresponding prototype verification system is designed to perform experimental tests in practical environments.
The carrier frequency is set as 3.5GHz, and the RIS array possesses 16$\times$16 units, each of which adopts two-bit phase quantization.
In particular, we adopt a self-developed broadband intelligent communication system 40MHz-Net (BICT-40N) terminal in order to fully acquire the channel information.
The terminal mainly includes a baseband board and a radio frequency (RF) front-end board, where the latter can achieve 26 dB transmitting link gain and 33 dB receiving link gain.
The orthogonal frequency division multiplexing (OFDM) signal is used for the terminal, where the bandwidth is 40MHz and the subcarrier spacing is 625KHz.
Also, the terminal supports a series of modulation modes, including QPSK, QAM, etc.
Through experimental tests, we validate a few functions and properties of the RIS as follows.
First, we validate a novel RIS power consumption model, which considers both the static and the dynamic power consumption.
As the RIS codebook switching frequency increases 1KHz, the dynamic power consumption of the whole RIS array will increase about 1mW.
Besides, we demonstrate the existence of the imaging interference and find that two-bit RIS can lower the imaging interference about 10 dBm.
Moreover, we verify that the RIS can outperform the metal plate in terms of the beam focusing performance.
Specifically, the received power at the desired focus point can be improved about 20 dBm with the aid of the RIS.
In addition, we find that the RIS has the ability to improve the channel stationarity.
Then, we realize the multi-beam reflection of the RIS utilizing the pattern addition (PA) algorithm.
Lastly, we validate the existence of the mutual coupling between different RIS units, which results in a maximum of 4 dB power gain.

\end{abstract}

\maketitle
\thispagestyle{empty}
\vspace{-1mm}



\section{Introduction}
\label{introduction}
Reconfigurable intelligent surface (RIS) has been proposed as a promising technology for the future wireless communication systems, due to its low power consumption, ease of deployment and flexible control of the incident signal \cite{1,2,3,4,30, 31,32}.
Specifically, the RIS is a planar metasurface consisting of a large number of passive units, each of which can be digitally controlled to induce an independent phase shift to the incident signal.
A specific phase shift pattern of the RIS is called a codeword, and all the codewords together form the RIS codebook \cite{100}.

To generate multiple different codewords, the RIS unit is mainly equipped with the PIN Diode.
As for the PIN Diode-based RIS, the authors in \cite{7,8} proposed an explicit power consumption model, which has been validated through experimental tests.
Specifically, both the static power consumption and the dynamic power consumption are taken into account, where the former is related to the control circuit of the RIS while the latter is mainly related to the polarization mode and the working status of the RIS.
However, in some scenes such as beam management \cite{9}, beam tracking \cite{10} and space-time modulated metasurfaces (STMM) \cite{200}, the RIS needs to perform codebook switching, i.e., the RIS codeword will vary with time in order to determine the position of the receiver or track the receiver.
It is worth noting that a rapid codebook switching frequency may also result in non-negligible dynamic power consumption.

With respect to the PIN Diode-based RIS, another important topic is the image interference due to the phase quantization, i.e., the phase of the RIS unit can only take a finite number of discrete values \cite{14}.
Note that existing works about RIS mainly consider the performance of the target receiver. 
For example, the authors in \cite{15} proposed that one-bit phase quantization was enough, due to the fact that the quantization bit number will have little impact on the performance of the target receiver.
Considering that there may exists many receivers in a practical communication system, we can not only focus on the performance of the target receiver, since the receiver located at the mirror image of the target receiver may be interfered and thus suffer from performance loss.
Besides, since the image interference is caused by the phase quantization, the quantization bit number may have a significant impact on the level of the image interference, which has not been discussed in existing works.

Moreover, the RIS and the metal plate possess some similarities, since they both can reflect the incident signal \cite{5}.
Many existing works \cite{6,11,12,13,21,22} discuss the advantages of RIS in terms of the beam coverage performance, where the angle power spectrum of the reflected signal was studied.
By comparison, the main advantage of the RIS over the metal plate lies in that the former can reflect the incident signal towards arbitrary direction via proper codebook design, while the latter can only realize specular reflection \cite{11}.
In addition, the channel stationarity should also be taken into account when making comparison between RIS and the metal plate \cite{23}.
Specifically, the equivalent channel along the path from the transmitter to the RIS and then to the receiver is modeled as the cascading convolution of three terms, including the transmitter-to-RIS channel, the RIS codeword and the RIS-to-receiver channel \cite{26}.
Thus, the RIS will have an impact on the stationarity of this equivalent channel.
On the one hand, if the RIS's codeword is selected to focus the incident signal towards the target receiver precisely, then the channel stationarity may be improved compared to case where the metal plate is deployed.
On the other hand, if the RIS performs codebook switching, then the channel stationarity may depend on the codebook switching frequency.
For example, a high switching frequency may induce non-stationarity.
Thus, the impact of the RIS on the channel stationarity also deserves investigating.

Note that the design criterion of the RIS codebook presents diversity.
Specifically, in addition to reflecting the incident signal towards a specific direction, the RIS can reflect the incident signal towards multiple directions simultaneously \cite{14,16,17,18}.
To achieve this, the authors in \cite{14,16,17,18} proposed the multi-beam codebook design algorithm.
However, they only provides the numerical simulation, which lacks the experimental tests to further verify the feasibility of the proposed algorithm.
Lastly, the potential mutual coupling among different RIS units should be considered since it exists inherently \cite{38,39}.
In \cite{19}, the authors derived a theoretical formula about the mutual coupling among different RIS units, while it lacks experimental tests to evaluate the effect of the mutual coupling on the received power.

With the above considerations, we propose a two-bit RIS-aided communication system and further design a corresponding prototype verification system to perform experimental tests and thus cover the shortages mentioned above within the existing works.
The carrier frequency is set as 3.5GHz, and the RIS array possesses 16$\times$16 units, each of which is of equal size.
The main characteristic of our prototype verification system lies in that we adopt a self-developed broadband intelligent communication system 40MHz-Net (BICT-40N) terminal in order to fully acquire the channel information.
The terminal mainly includes a baseband board and a radio frequency (RF) front-end board, where the latter can achieve 26 dB transmitting link gain and 33 dB receiving link gain.
The orthogonal frequency division multiplexing (OFDM) signal is used for the terminal, where the bandwidth is 40MHz and the subcarrier spacing is 625KHz.
Also, the terminal supports a series of modulation modes, including QPSK, QAM, etc.
The main contributions of this paper are summarized as follows:
\begin{itemize}
   \item We propose a new power consumption model of the RIS, which considers both the static power consumption due to the hardware structure and the dynamic power consumption due to the codebook switching.
We find that as the RIS codebook switching frequency increases 1KHz, the dynamic power consumption of the whole RIS array will increase about 1mW, which can not be neglected.
     \item We test the differences between the RIS and the metal plate from two perspectives.
   For one thing, the RIS can outperform the metal plate in terms of the beam focusing performance.
   Specifically, with the aid of the RIS, the received power at the desired focus point is about 20 dBm higher than that at the surrounding points.
   For another, we find that if the RIS is designed to reflect the incident signal towards the receiver, then it can significantly improve the channel stationarity compared to the metal plate.
   However, if the RIS is performing codebook switching and its reflected beam scans a range centered on the receiver, then the channel stationary will depend on the codebook switching frequency.
 \item We validate the existence of the image interference at the mirror image of the target receiver, and further test the impact of the quantization bit number on the image interference.
  Specifically, the image interference under two-bit phase quantization is about 10 dBm lower than that under one-bit phase quantization, which is a significant finding.
  \item We test the feasibility of the RIS-aided multi-beam reflection.
  Specifically, we consider the dual-beam reflection and set the beam resolution as about $8^{\circ}$.
  We find that the RIS succeeds in realizing multi-beam reflection separately towards $\{(10^{\circ}, 0^{\circ}), (10^{\circ}, 180^{\circ})\}$, $\{(15^{\circ}, 0^{\circ}), (15^{\circ}, 180^{\circ})\}$ and $\{(30^{\circ}, 0^{\circ}), (30^{\circ}, 180^{\circ})\}$, while it fails to realize multi-beam reflection towards $\{(5^{\circ}, 0^{\circ}), (5^{\circ}, 180^{\circ})\}$.
  \item We verify the existence of the mutual coupling between different RIS units.
  Specifically, we compare the practical measured received power with its theoretical value without the effect of the mutual coupling and find a gap of 4 dB between them.
  Considering that we have excluded the effect of the surrounding environment, this gap can only be due to the effect of the mutual coupling.
  Moreover, we find that as the number of the invalid RIS units increases, the gap increases accordingly.
\end{itemize}

The rest of the paper is organized as follows.
In Section II, we introduce the system model of the RIS-aided system.
In Section III, we display the test platform.
The experimental results are shown in Section IV and the conclusion is given in Section V.

Notations: We use the following notation throughout this paper:
Vectors and matrices are denoted in bold font;
$\mathbf A^{-1}$ denotes the inversion matrix of $\mathbf A$;
$\mathbf a^{T}$ denotes the transpose of $\mathbf a$;
$\mathbb C$ denotes the complex number sets;
$\mathbb E[\cdot]$ denotes the expectation operation;
$|\cdot|$ and $\angle(\cdot)$ separately mean extracting the amplitude and the phase of a complex number, and $j=\sqrt{-1}$ is the imaginary unit;
$v\sim \mathcal {CN}(0, \sigma^2)$ means that $v$ follows the complex Gaussian distribution with zero-mean and variance $\sigma^2$.
Moreover, diag$(\mathbf a)$ is a diagonal matrix whose diagonal elements are  the elements of $\mathbf a$.
\section{System Model}
\subsection{Signal Model}
\begin{figure}[!t]
	\centering
	\includegraphics[width=105mm]{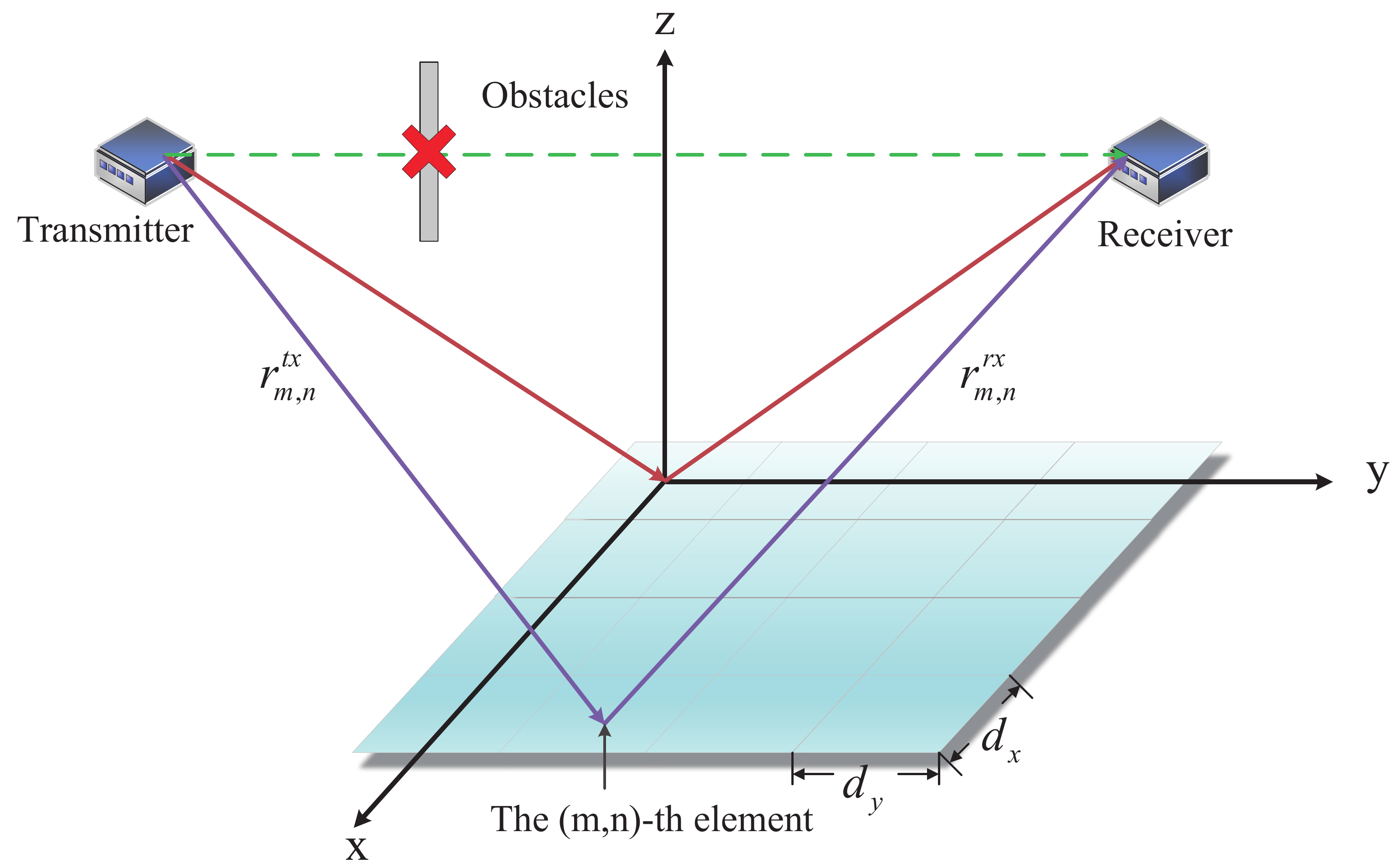}
	\caption{A RIS-aided wireless communication system.}
    \label{scene}
\end{figure}

We consider a RIS-aided wireless communication system, where a RIS is deployed to aid the communication between a single-antenna transmitter and a single-antenna receiver, as illustrated in Fig. \ref{scene}.
We assume that the direct path between the transmitter and the receiver is blocked completely.
The RIS is placed in the xy-plane and has a uniform planar array structure of $M\times N$ units, each of which is of equal size $d_x\times d_y$.
Here, we have $d_x= d_y= \frac{c}{2f_0}$, where $f_0$ and $c$ denote the carrier frequency and the speed of light, respectively.
We assume that the amplitude coefficient of all the RIS units is equal to 1, and we denote the phase shift of the $(m, n)$-th RIS unit at time $t$ as $\varphi_{m,n}[t]$, where $m=1, 2,...,M$ and $n=1, 2,...,N$.
Specifically, $\varphi_{m,n}[t]\in \mathcal V = \{0, \frac{2\pi}{K}, ..., \frac{(K-1)2\pi}{K}\}$, where $K=2^b$ denotes the quantization level and $b$ denotes the quantization bit number.
Moreover, let us define $\mathbf a[t]=(e^{j\varphi_{1,1}[t]}, e^{j\varphi_{1,2}[t]},...,e^{j\varphi_{M, N}[t]})^T\in \mathbb C^{MN\times 1}$ and $\mathbf \Theta[t]$ = diag($\mathbf a[t]$), where $\mathbf a[t]$ represents the RIS codeword at time $t$ and is selected from the predesigned RIS codebook.
In the following, we refer to the variation of $\mathbf a[t]$ with time $t$ as {\it codebook switching} or {\it beam switching}.
The channel from the transmitter and the receiver to the RIS at time $t$ can be defined as $\mathbf h_{tx}[t]\in \mathbb C^{MN\times 1}$ and $\mathbf h_{rx}[t]\in \mathbb C^{MN\times 1}$, respectively.
Then, the received signal at the receiver at time $t$ can be expressed as
\begin{align}\label{channel}
         r[t]= \mathbf h^{T}_{rx}[t]\mathbf \Theta[t]\mathbf h_{tx}[t]s[t]+w[t],
\end{align}
where $s[t]$ and $w[t]\sim \mathcal {CN}(0, \sigma^2)$ separately denote the transmitted signal and the additive white Gaussian noise (AWGN).
\subsection{Channel Stationarity Model and the Received Power Model}
Let us define the stationarity of the end-to-end equivalent RIS channel.
At time $t$, the equivalent RIS channel is denoted as $c[t] = \mathbf h^{T}_{rx}[t]\mathbf \Theta[t]\mathbf h_{tx}[t] $.
The statistic characteristics of $c[t]$ include the channel mean $a[t]= \mathbb E\{c[t]\}$, and the covariance coefficient $R[t_0,t] = \mathbb E\{c[t_0]c^{\ast}[t-t_0]\}$.
Since $R[t_0,t]$ is related to the initial time $t_0$ and the time interval $\Delta_t = t-t_0$, thus $R[t_0,t]$ can be rewritten as $R[t_0,\Delta_t]$.
According to the generalized definition of stationarity for a stochastic process, if $a[t]$ is constant and $R[t_0,\Delta t]$ is independent on $t_0$, then $c[t]$ is stationary.
However, this definition of the stationary is hard to evaluate in practical tests.
Thus, we use the following criterion to evaluate the channel stationary in the rest of this paper.
Let us define the normalization difference of the RIS channel's statistic characteristics between time $t$ and the initial time $0$ as
\begin{align}\label{delta}
\delta_x[t] = \frac{|x[t]-x[0]|^2}{|x[0]|^2},
\end{align}
where $x[t]$ can be substituted by $a[t]$ or $R[t_0,t]$.
If $\delta_x[t]$ is less than the given threshold in the time duration $[0,T]$, then $c[t]$ in $[0,T]$ is considered stationary.

The received power $P_r[t]$ of the receiver at time $t$ can be computed as $|r[t]|^2$.
Specifically, as for the case where both $\mathbf h_{tx}[t]$ and $\mathbf h_{rx}[t]$ experience the line-of-sight (LoS) path fading, the authors in \cite{11} have introduced a more accurate analytical model of $P_r[t]$, which neglects the noise's effect and can be formulated as
\begin{align}\label{Pr}
         P_r[t]= P_t\frac{G_tG_rGd_xd_y\lambda^2}{64\pi^3}\times\left|\sum_{m=1}^M\sum_{n=1}^N\frac{1}{r_{m,n}^{tx}[t]r_{m,n}^{rx}[t]}e^{j\varphi_{m, n}+\frac{-j2\pi(r_{m,n}^{tx}[t]+r_{m,n}^{rx}[t])}{\lambda}}\right|^2,
\end{align}
where $P_t$ represents the transmitted power, $G_t$, $G_r$ and $G$ separately represent the antenna gain of the transmitter, the receiver and the RIS, $r_{m,n}^{tx}[t]$ and $r_{m,n}^{rx}[t]$ separately denotes the distance from the transmitter and the receiver to the $(m, n)$-th RIS unit at time $t$, and $\lambda=\frac{c}{f_0}$ is the wavelength.
Note that \eqref{Pr} is proved to be reasonable through measurement-based validations and the radiation pattern of the RIS unit is assumed to be isotropic.
\subsection{RIS Codebook Design Criterion}
It can be seen from \eqref{Pr} that $\varphi_{m, n}[t]$ will influence the received power.
Thus, it should be designed properly.
Specifically, if we want to reflect a incident signal with incident angle ($\theta_i$, $\phi_{i}$) towards the target exit angle ($\theta_r$, $\phi_{r}$), then $\varphi_{m, n}[t]$ should be designed as \cite{11}
\begin{align}\label{singlebeam}
	\varphi_{m, n}[t]={\rm mod}\left(-\frac{2\pi}{\lambda}((\Psi_{i}+\Psi_{r})(m-0.5)d_x+(\Phi_{i}+\Phi_{r})(n-0.5)d_y), 2\pi\right),
\end{align}
where we define $\Psi_{i}=\sin\theta_{i}\cos\phi_{i}$, $\Psi_{r}=\sin\theta_{r}\cos\phi_{r}$, $\Phi_{i}=\sin\theta_{i}\sin\phi_{i}$ and $\Phi_{r}=\sin\theta_{r}\sin\phi_{r}$ for ease of notation and \eqref{singlebeam} is the well-known {\it generalized Snell's law of reflection}.
Since $\varphi_{m, n}[t]$ derived from \eqref{singlebeam} must be quantified to take value from $\mathcal V$, the imaging interference will emerge \cite{14}, which means that the RIS will also reflect the incident signal towards ($\theta_r$, $\phi_{r}+\pi$).
Note that this reflected beam towards ($\theta_r$, $\phi_{r}+\pi$) is inherent and undesirable.

Moreover, if we want to realize multi-beam reflection, i.e., to reflect a incident signal with incident angle ($\theta_i$, $\phi_{i}$) to a total of $R$ target exit angles denoted as ($\theta_1$, $\phi_1$), ($\theta_2$, $\phi_2$),..., ($\theta_R$, $\phi_R$), the pattern addition (PA) technique \cite{14} should be adopted.
Specifically, $\varphi_{m,n}[t]$ in this case should be designed as
\begin{align}\label{multibeam}
         \varphi_{m,n}[t]=\angle\left(\sum_{r=1}^Rw_re^{j\varphi_{m,n}^r[t]}\right),
\end{align}
where $w_r$ is the weight of the $r$-th reflected beam, and $\varphi_{m,n}^r[t]$ is designed according to \eqref{singlebeam}, which produces the $r$-th reflected beam towards ($\theta_r$, $\phi_{r}$), $r=1,2,...,R$.
Likewise, the imaging interference also exists for the multi-beam reflection, i.e., the RIS will also focus the incident signal towards
($\theta_1$, $\phi_1+\pi$), ($\theta_2$, $\phi_2+\pi$),..., ($\theta_R$, $\phi_R+\pi$).

\section{Test Platform}
\subsection{The Overall Schematic Diagram}
\begin{figure}[!t]
	\centering
	\subfigbottomskip=2pt
	\subfigcapskip=-5pt
	\subfigure[]{
		\includegraphics[width=0.42\linewidth]{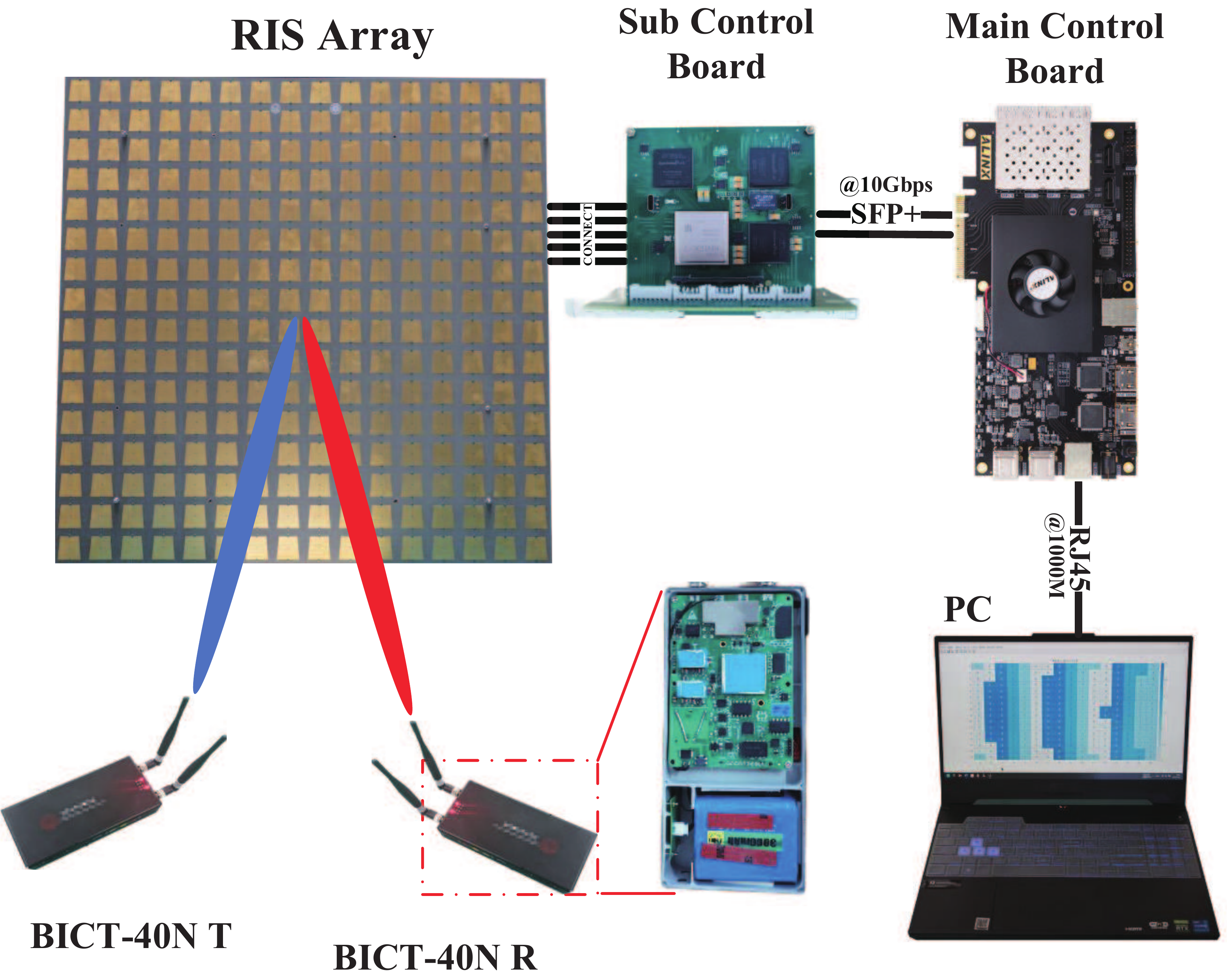}}
	\subfigure[]{
		\includegraphics[width=0.47\linewidth]{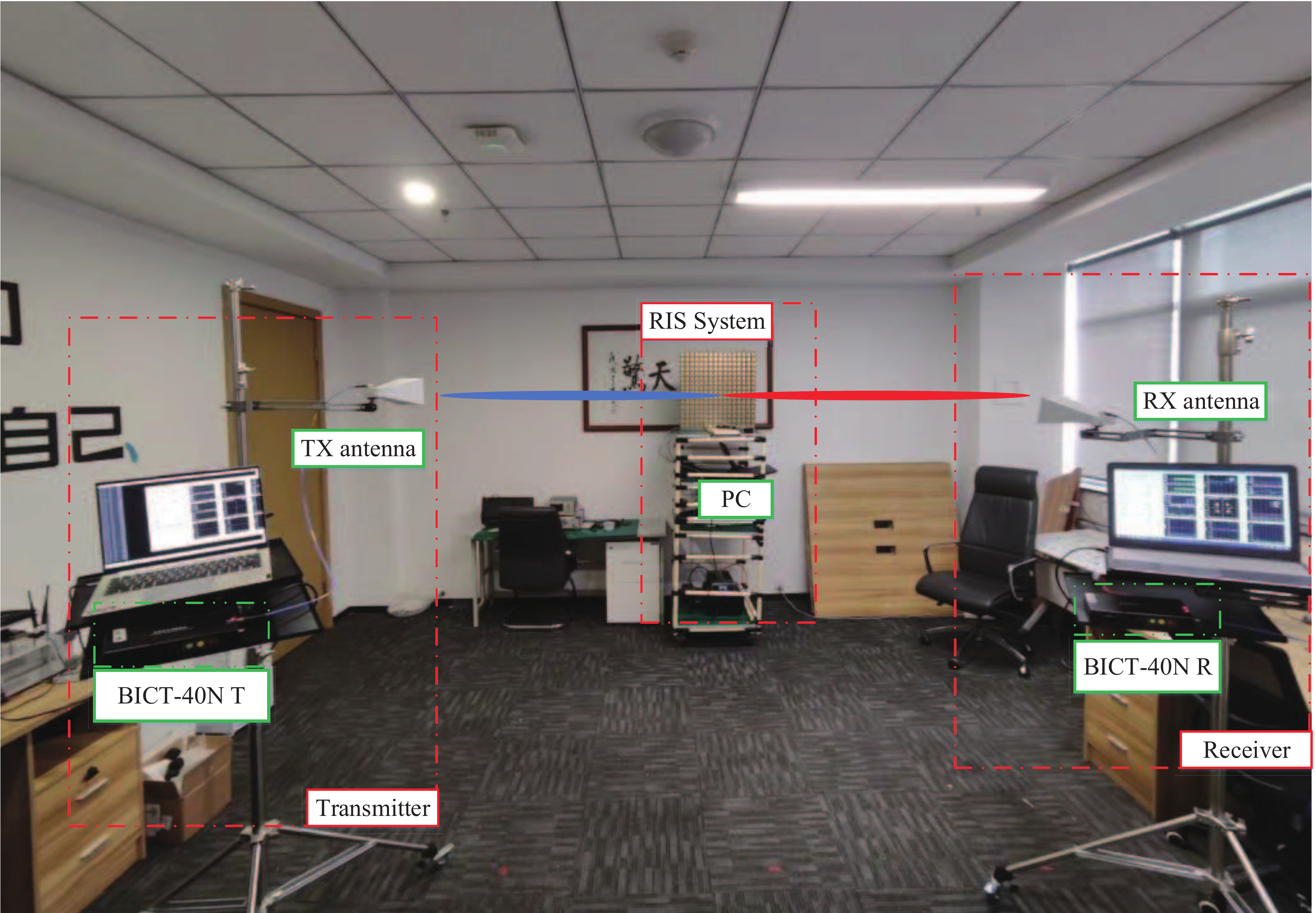}}
\caption{The proposed two-bit RIS aided communication system. (a) Schematic diagram. (b) Practical test environment.}
\label{diagram}
\end{figure}

The overall schematic diagram of the proposed two-bit RIS aided communication system is shown in Fig. \ref{diagram}(a), which includes a two-bit RIS and a BICT-40N terminal.
Specifically, the two-bit RIS consists of a RIS array, a sub-control board, a main control board and a personal computer (PC).
The PC is responsible for the calculation and generation of the RIS codebook.
After the codebook is generated, it is sent to the main controller through user datagram protocol (UDP) protocol.
Then, the main control board separates the command information from the codebook information, and distributes them to the sub-control board through the Small Form-factor Pluggable (SFP)+ interface.
The sub-control board is connected to RIS array through high-density connector and can realize independent high-speed control of the RIS unit.
Note that BICT-40N is a full-stack self-developed terminal, which supports point-to-point communication based on 802.11n standard.
Unlike the traditional systems, the BICT-40N terminal is able to collect and upload specific data from the physical layer (PHY) in real time, such as the channel state, the frequency deviation estimation and the constellation diagram.
Thus, the BICT-40N terminal is capable of performing the data acquisition tasks required by the system.
In Fig. \ref{diagram}(b), we display a practical test environment with respect to the proposed two-bit RIS aided communication system.
\subsection{The RIS Unit Structire and the RIS control circuit}

\begin{figure}[!t]
	\centering
	\subfigbottomskip=2pt
	\subfigcapskip=-5pt
	\subfigure[]{
		\includegraphics[width=0.5\linewidth]{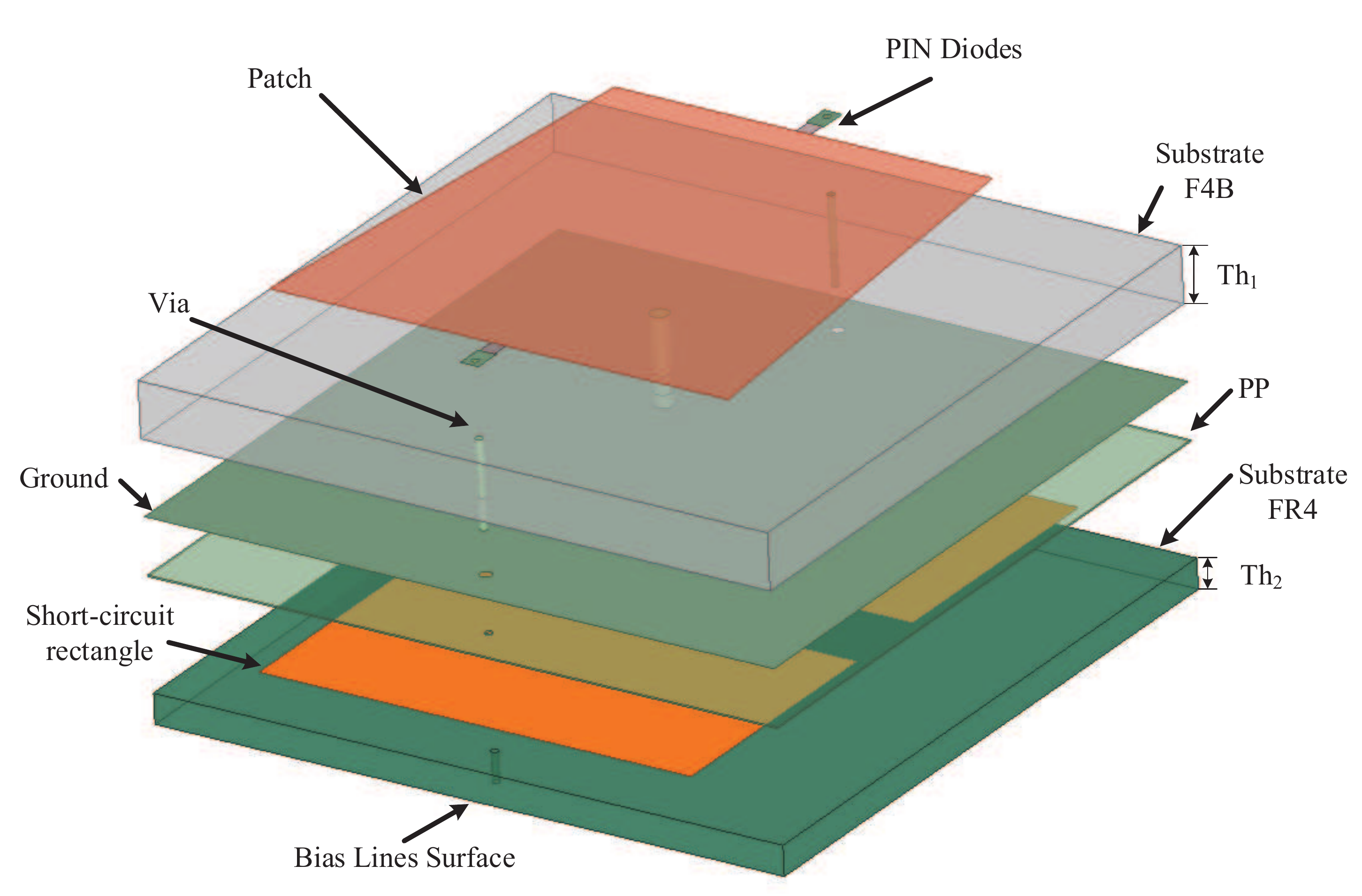}}
	\subfigure[]{
		\includegraphics[width=0.45\linewidth]{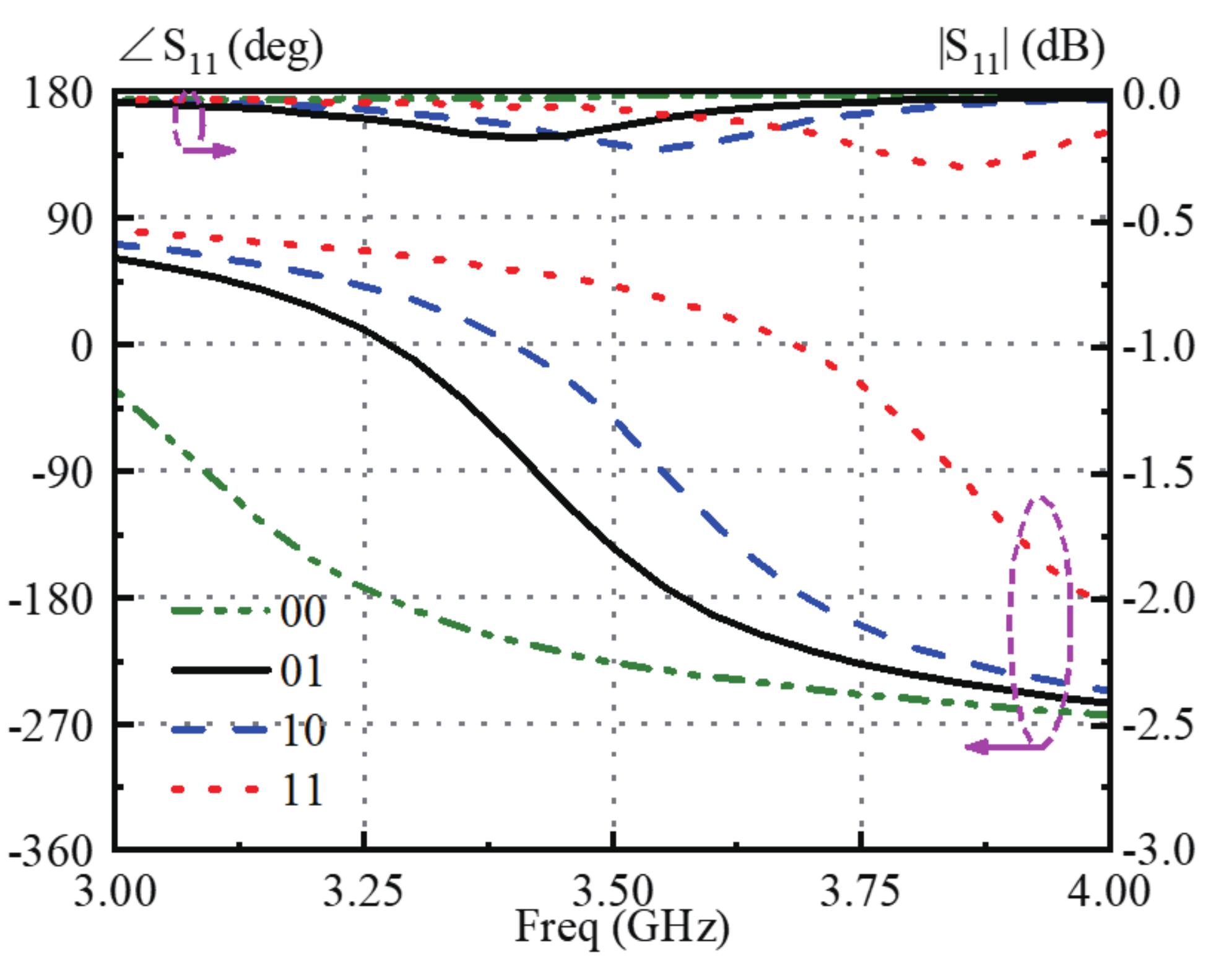}}
\caption{(a) The RIS unit structure. (b) The RIS unit's amplitude/phase response.}
\label{RIS structure}
\end{figure}

\begin{figure}[!t]
	\centering
	\subfigbottomskip=2pt
	\subfigcapskip=-5pt
	\subfigure[]{
		\includegraphics[width=0.38\linewidth]{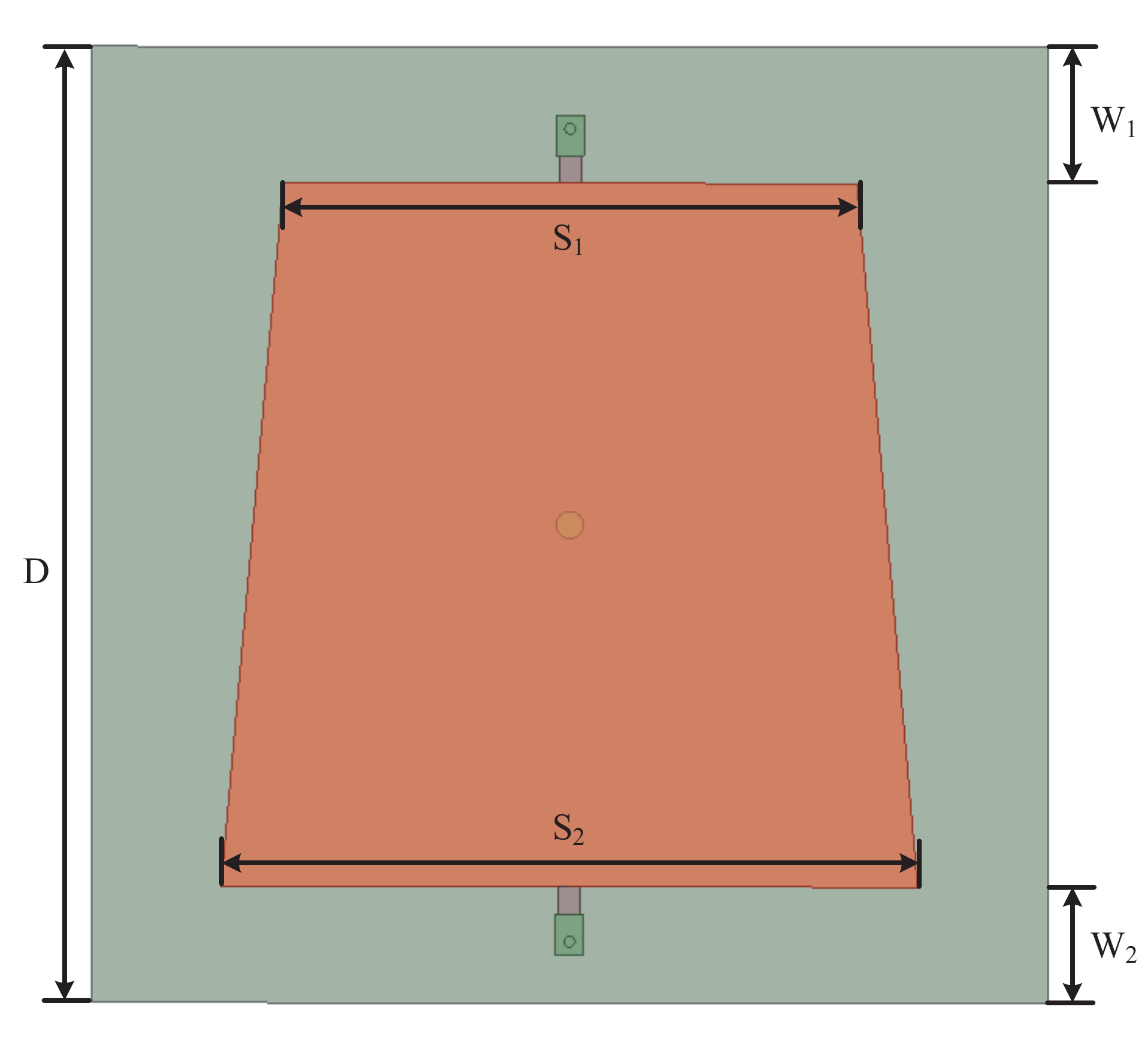}}
    \subfigure[]{
        \includegraphics[width=0.35\linewidth]{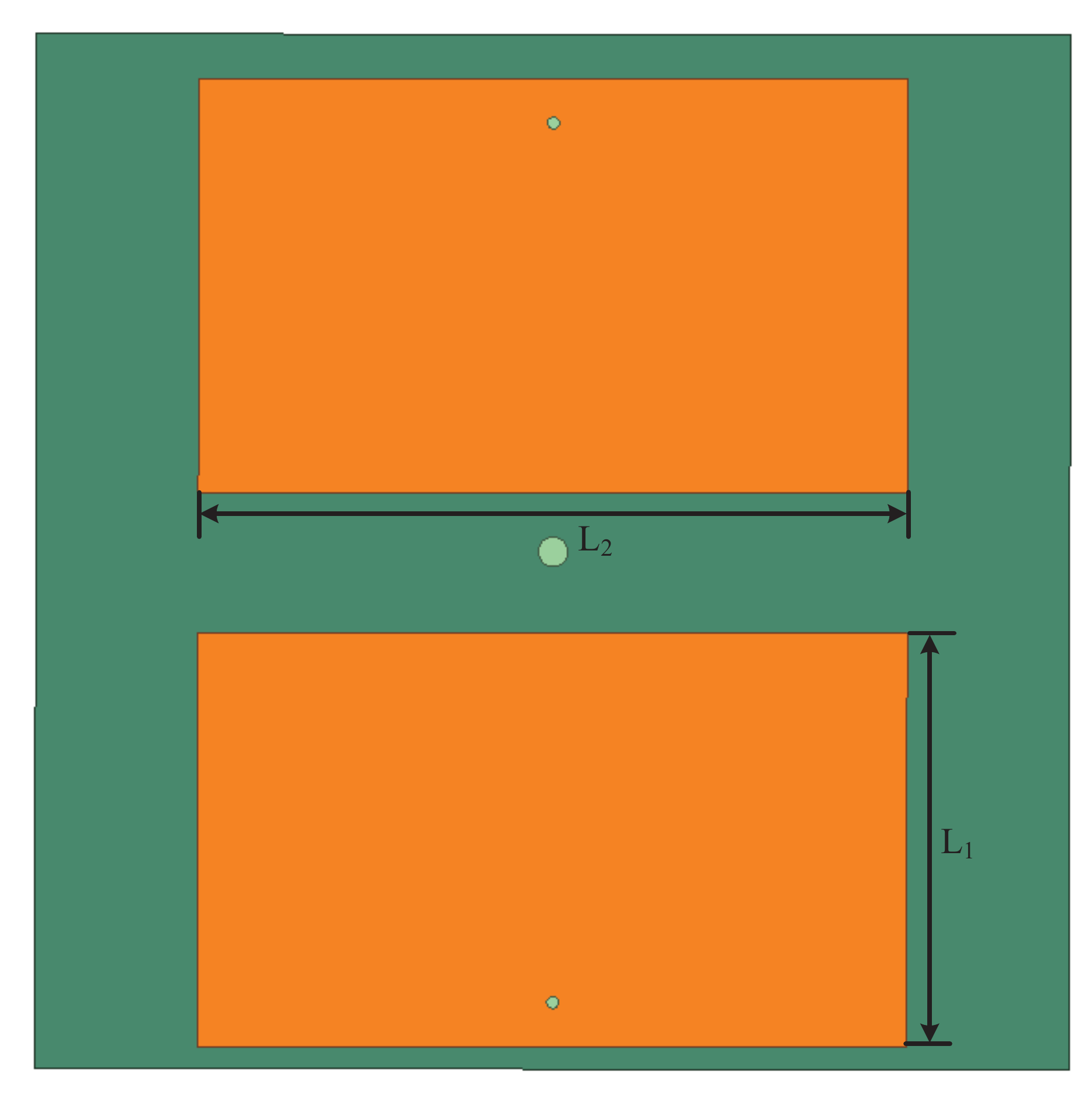}}
\caption{(a) The top layer's patch antenna. (b) The short-circuit rectangle.}
\label{structure11}
\end{figure}

The structure and the amplitude/phase response of the RIS unit are shown in Fig. \ref{RIS structure}(a) and Fig. \ref{RIS structure}(b), respectively.
Specifically, the top layer of the RIS unit is equipped with a microwave patch antenna, which is connected to the bias line through two PIN Diodes and metallized vias.
The design of the microwave patch antenna is shown in Fig. \ref{structure11}(a), where $\text {D}$ = 35mm, $\text {W}_1$ = 5mm, $\text {W}_2$ = 4.25mm, $\text {S}_1$ = 21mm and $\text {S}_2$ = 25.5mm.
The metallized via's diameter is set as 0.4mm.
The PIN Diode used here is SMP1345-079LF (Skyworks), which possesses small reverse capacitance and can provide good forward and reverse characteristics at sub-6GHz.
The substrate material for both the top layer and layer1 is F4B (dielectric constant  $\varepsilon_r$ = 2.65, dielectric loss angle tangent tan$\delta$ = 0.001) with a thickness of $\text {Th}_1$ = 3mm.
Moreover, layer2 is the short-circuit rectangle, and the material used between layer1 and layer2 is Shengyi 1080pp ($\varepsilon_r$ = 3.67) with a thickness of 0.08mm.
Fig. \ref{structure11}(b) displays the design of the short-circuit rectangle, where $\text {L}_1$ = 14mm and $\text {L}_2$ = 24mm.
The bottom layer is the bias lines surface, and the substrate material of both layer2 and the bottom layer is FR4 ($\varepsilon_r$ = 4.2, tan$\delta$ = 0.02) with a thickness of $\text {Th}_2$ = 1.6mm.
The amplitude response $|S_{11}|$ and the phase response $\angle S_{11}$ of the RIS unit versus the frequency were simulated by ANSYS HFSS.
We can see from Fig. \ref{RIS structure}(b) that the RIS unit can realize two-bit phase control and the phase spacing is close to $90^{\circ}$ within the frequency band from 3.4GHz to 3.6GHz.

\begin{figure}[!t]
	\centering
	\subfigbottomskip=2pt
	\subfigcapskip=-5pt
	\subfigure[]{
		\includegraphics[width=0.43\linewidth]{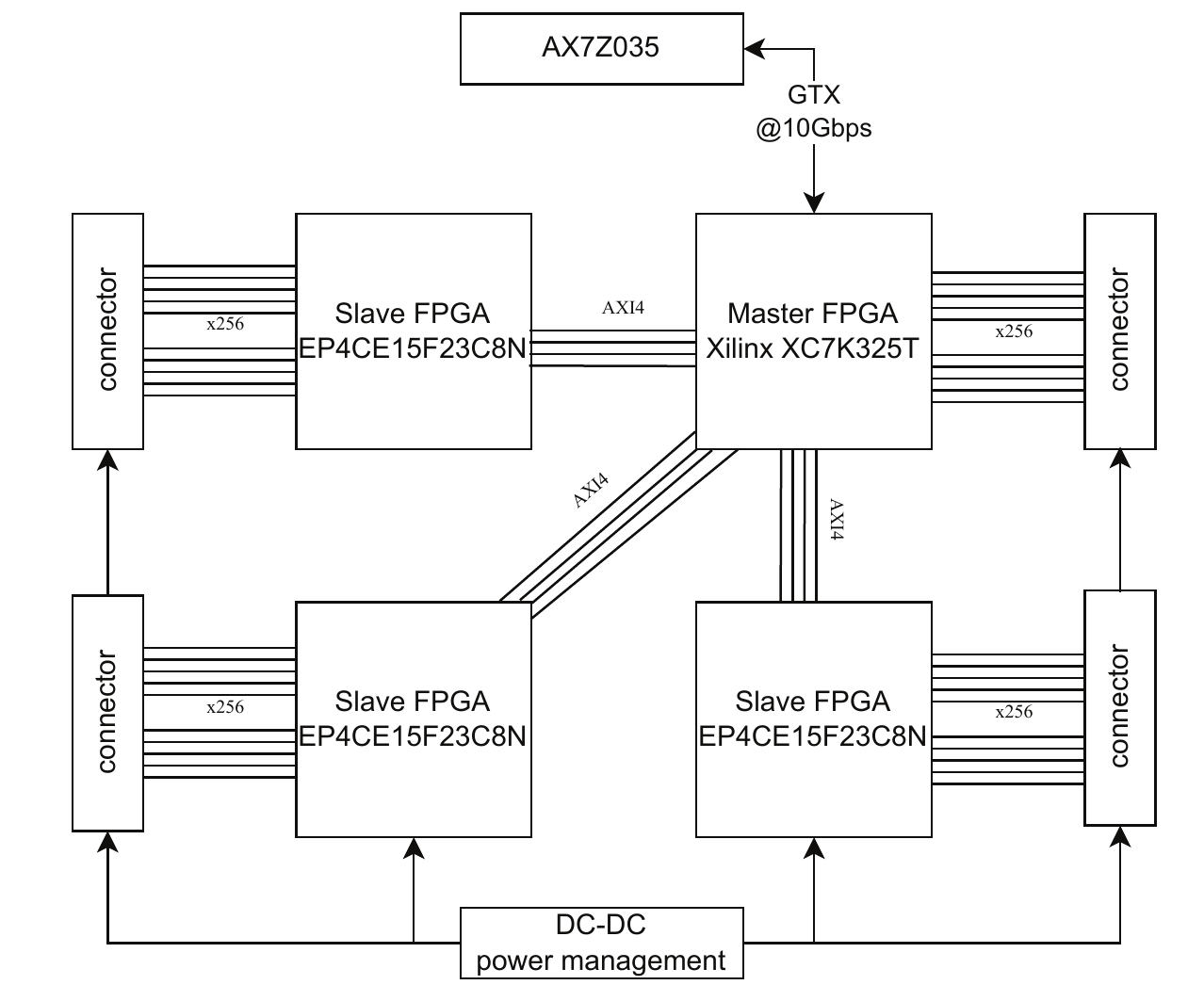}}
	\subfigure[]{
		\includegraphics[width=0.41\linewidth]{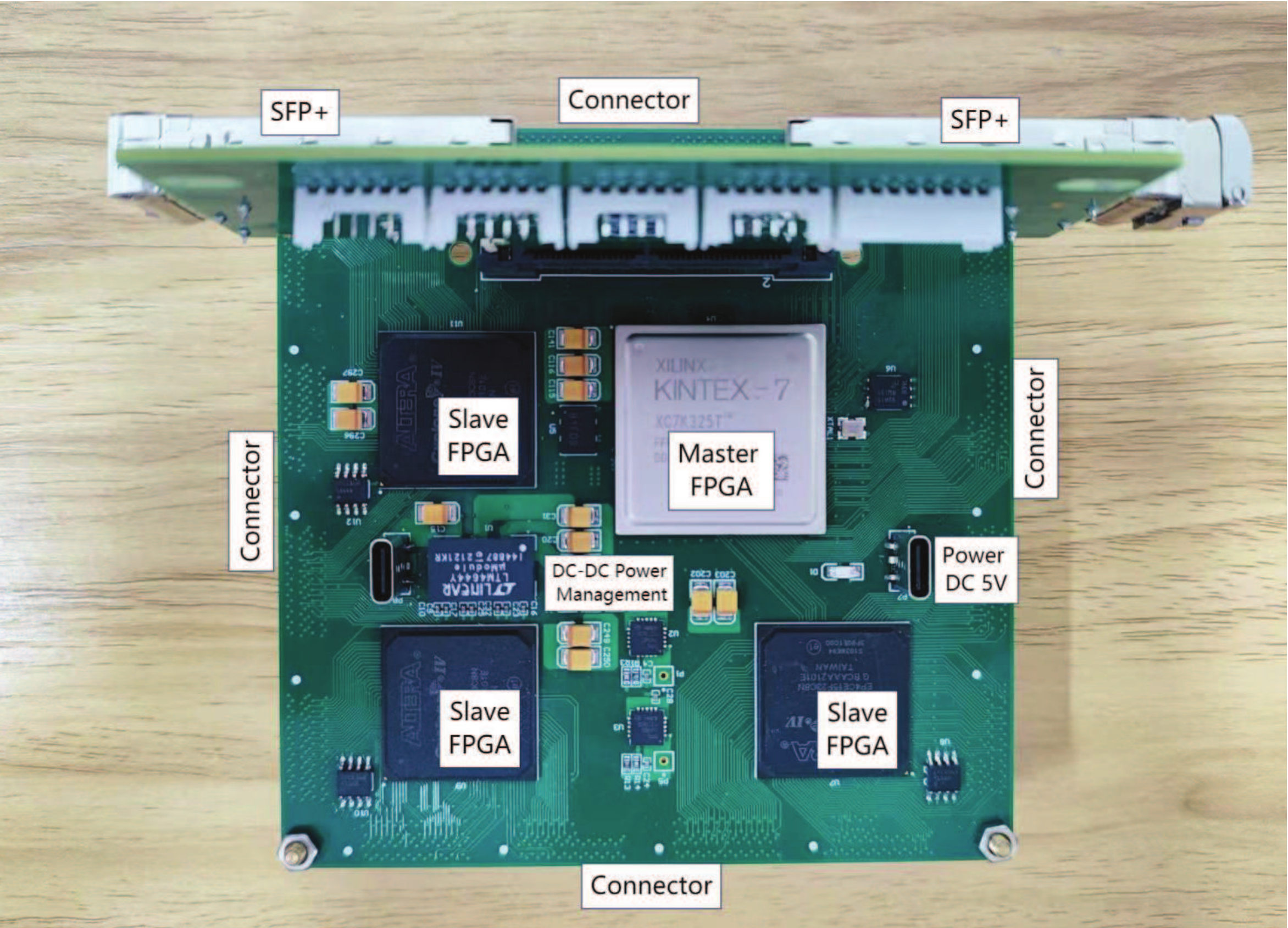}}
\caption{The RIS control circuit. (a) Schematic diagram. (b) Physical diagram.}
\label{control}
\end{figure}

To facilitate the channel stationarity test, the RIS needs a relative high codebook switching frequency, which places high demands on the controller's switching frequency and delay.
Thus, we use three low-cost, high I/O density Cyclone IV EP4CE15F23C8N field programmable gate arrays (FPGAs) cascaded with a high-performance Xilinx XC7K325T FPGA, which can run at 300MHz system clock frequency and control each PIN Diode individually.
The schematic diagram and the physical diagram of the RIS control circuit are shown in Fig. \ref{control}.
The circuit includes a power management part, a bias circuit, a master FPGA controller, three slave FPGA controllers and an external GTX communication interface.
The master FPGA is responsible for controlling the 256 PIN Diodes and the external GTX cascade codebook refresh, and the three slave FPGAs control the remaining 768 PIN diodes.
The master FPGA and the slave FPGAs are connected via a parallel 32-bit AXI-Stream4 bus, whose signal lines consist of 32-bit-tdata, 1-bit-tready, 1-bit-tvalid and 1-bit-tclk.
The bus follows the AXI4-Stream protocol for communication with a reference clock frequency of 100MHz.
Besides, it enables high-speed data transfer in the direction from the master to the slave with a peak throughput rate of 3.2Gbps.
\subsection{The BICT-40N Terminal}
\begin{figure}[!t]
	\centering
	\includegraphics[width=100mm]{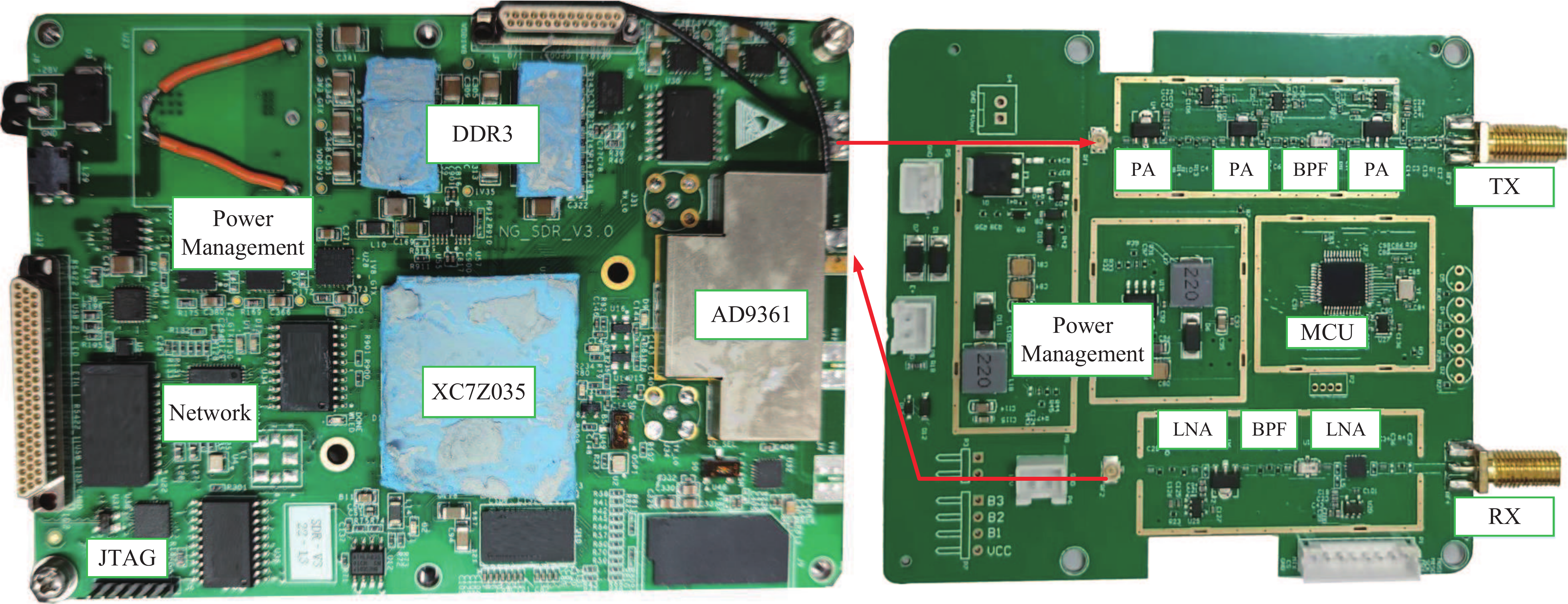}
	\caption{{The physical diagram of the BICT-40N terminal.}}
	\label{ternimal}
\end{figure}

The BICT-40N terminal is utilized for channel stationary tests.
As shown in Fig. \ref{ternimal}, the BICT-40N terminal mainly includes a baseband board and a RF front-end board, which are connected through the coaxial line and the power line.
The baseband board card contains a ZYNQ chip XC7Z035 and a RF transceiver chip AD9361, which is used for signal transceiving and processing.
Also, it can support the carrier frequency within the range from 70MHz to 6GHz and a maximum of 56MHz RF bandwidth.
Besides, through external synchronous clock source, the errors introduced by the carrier frequency offset (CFO) can be reduced significantly during the channel state measurement.
The RF front-end board uses Qorvo's TQP3M9009 and TQP7M9103 chips and Mini-Circuits's PMA3-83LN+ chip for the transceiver link.
In addition, the terminal can use a horn antenna, which can provide additional gain.
\subsection{Digital Signal Processing}

\begin{figure}[!t]
	\centering
	\includegraphics[width=75mm]{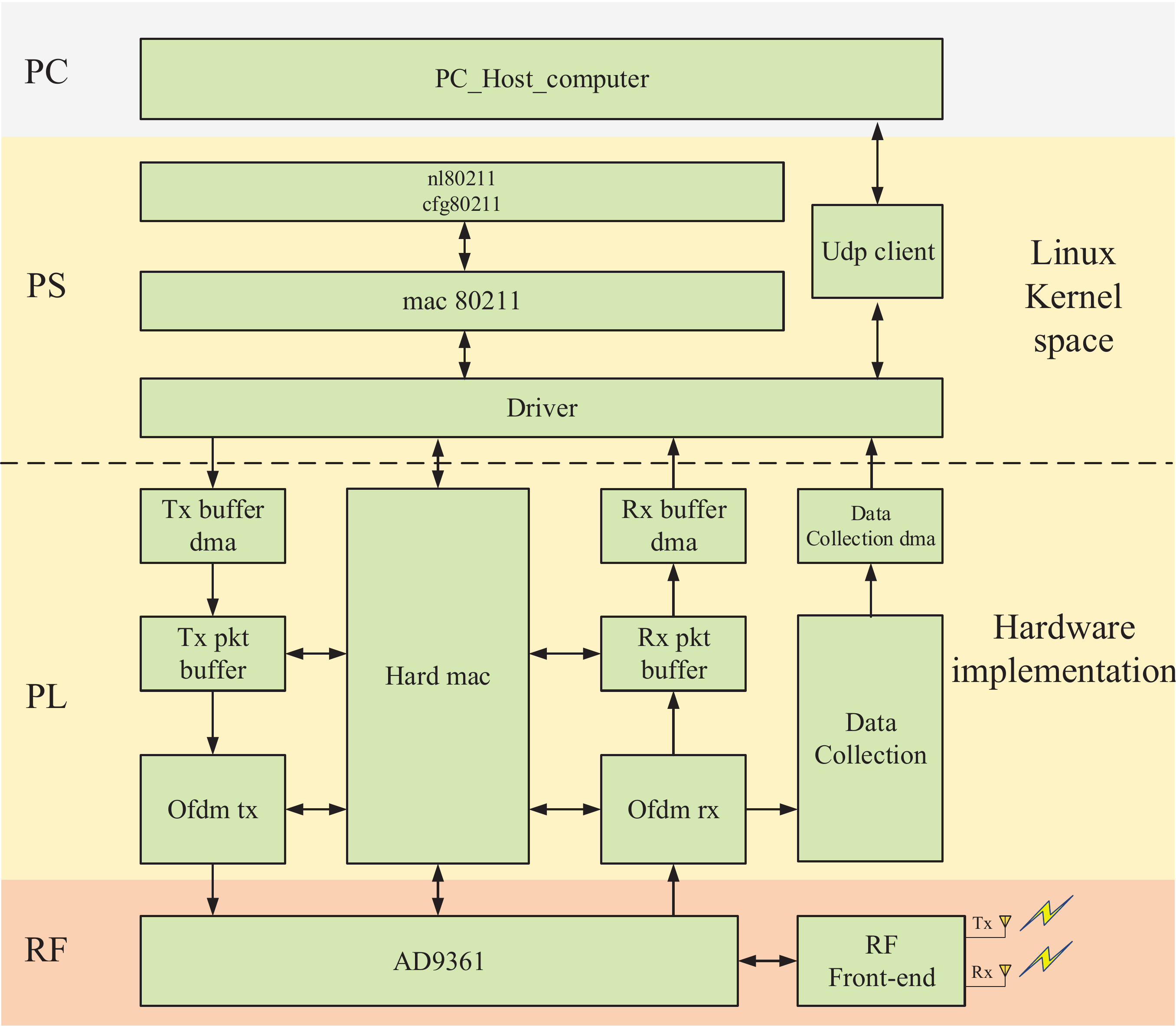}
	\caption{{The digital signal processing flow of the communication system.}}
	\label{diagram2}
\end{figure}

The digital signal processing of the system is mainly divided into two parts, including the transmitter part and the receiver part.
On the transmitter side, the PC upper computer transmits the service data to the ARM processor of XC7Z035 chip by UDP protocol.
The processor uses the LwIP protocol stack to parse the UDP messages.
Specifically, the data is unpacked and encapsulated into wireless data frames through mac80211 in the linux kernel, and is sent to the hard mac through AXIDMA.
Moreover, the hard mac runs in the programmable logic part of the chip and implements fallback, CRC check, ACK answer and retransmission operations in the MAC layer.
The data is then passed to the OFDM-based PHY layer, which is capable of encoding, modulating, and RF outputting the bitstream data from the MAC layer.
Similarly, the receiver can receive, synchronize, demodulate and decode the RF data, and finally send the data to the PC through UDP protocol.
Since it is self-developed, the system can collect and display the intermediate information at the receiver in real time.
After receiving the signal, the receiver first determines the data to be collected according to the frame format configured on the PS side.
Then, the corresponding data is obtained from the PHY layer for framing, and the data is carried to the PS side through the direct memory access (DMA) interface and sent to the upper computer by UDP protocol for real-time data display and storage.
Specifically, the collected information includes channel state information, phase tracking effect, constellation diagram, signal-to-noise ratio, RSSI, etc.
The digital signal processing flow is summarized in Fig. \ref{diagram2}.

The PHY layer algorithm of the system is based on the IEEE802.11n standard with a transmission bandwidth of 40MHz and 52 data subcarriers and four pilot subcarriers.
The transmitter of the PHY layer consists of seven operations, including frame formation, scrambling, convolutional coding, interleaving, QAM modulation, IFFT, and cyclic prefix (CP) insertion.
Similarly, the receiver's operation contains frame detection (short training sequence synchronization), timing synchronization (long training sequence synchronization), CFO Correction, FFT, channel estimation, demodulation, deinterleaving, data acquisition, decoding and descrambling.
\section{Experimental Results}
\subsection{Testing Devices and System Parameters}
The testing devices of the proposed prototyping system mainly include a transmitter, a receiver, a RIS and a PC.
Specifically, the transmitter consists of a RF signal source RIGOL-DSG3060 and a TX antenna, where the transmitted signal is produced by the RF signal source and then sent to the atmosphere through the TX antenna.
Moreover, the RIS is deployed to reflect the incident signal and the PC is utilized to calculate and generate the RIS codebook.
The receiver consists of a RX antenna and a spectrum analyzer Tektronix-RSA306, which are separately responsible for the reception and the post-processing of the received signal reflected by the RIS.
Note that both the transmitter and the receiver adopt the horn antenna.
Besides, for the channel stationary test, we adopt the BICT-40N terminal to acquire the channel information.
The BICT-40N terminal works at 3.5GHz with 40MHz bandwidth.
The OFDM signal is used for the terminal and the subcarrier spacing is set as 625KHz.

The system parameters need to be configured are summarized as follows.
The carrier frequency is set as 3.5GHz.
The RIS array possesses $16\times 16$ units, each of which adopts two-bit phase quantization.
The antenna heights of the transmitter and the receiver are separately denoted as $h_t=1.65$m and $h_r=1.8$m.
The height $h_s$ of the RIS center is set as 1.8m.
Besides, the antenna gain $G_t$ and $G_r$ are 13 dBi, and the transmitted power $P_t$ is 0 dBm.
The incident angle $(\theta_i,\phi_i)$ is set as $(0^{\circ}, 270^{\circ})$ while the target exit angle $(\theta_r, \phi_r)$ is set as $(30^{\circ}, 0^{\circ})$.
The distances from the transmitter and the receiver to the RIS center are separately denoted as $d_t=0.5$m and $d_r=3$m.
\subsection{The BICT-40N Terminal Performance Test}
\begin{figure}[!t]
	\centering
	\includegraphics[width=100mm]{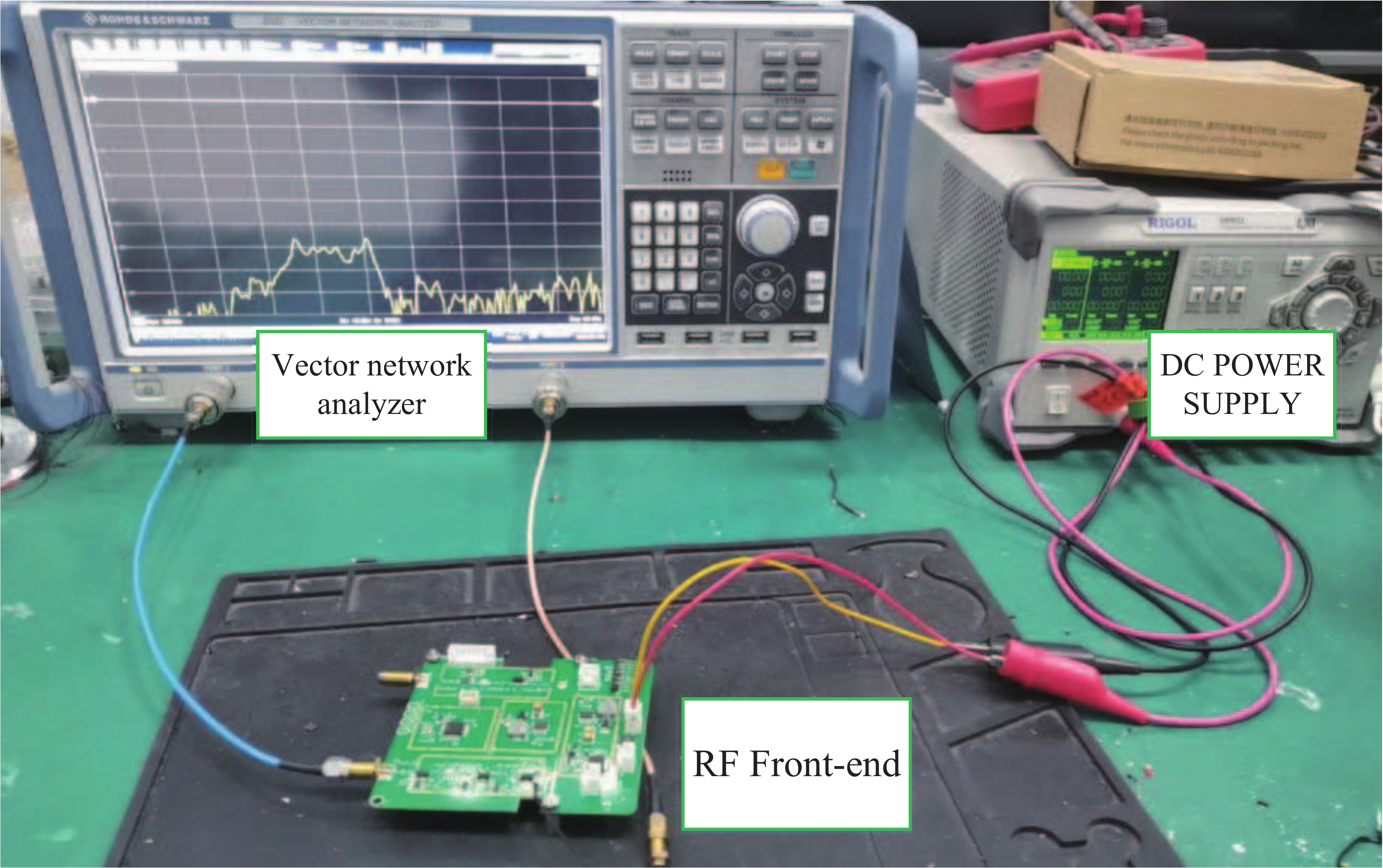}
	\caption{{Test environment of the RF front-end board.}}
	\label{terminal_RF_front}
\end{figure}
Tested by the spectrum analyzer Tektronix RSA306 and the RF signal source RIGOL DSG3060, the maximum output power of the chip AD9361 of the baseband board is 0 dBm and its receiving sensitivity is $-70$ dBm.
Besides, the RF front-end board is tested by the vector network analyzer R$\&$S ZEB, as shown in Fig. \ref{terminal_RF_front}.
The measured RF front-end link performance of the BICT-40N terminal is shown in Tab. \ref{parameter0}.
Note that the baseband board and RF front-end are cascaded to enable the BICT-40N terminal to achieve a maximum transmitted power of 25 dBm and a receiving sensitivity of $-83$ dBm, which meets the system's test requirements.

\begin{table}[]
\caption{The measured RF front-end link performance.}
\centering
\begin{tabular}{c c} \hline
	Parameters & Default values\\ \hline \hline
	Bandwidth  &40MHz\\ \hline
    Carrier frequency &3.5GHz\\ \hline
    Maximum transmitted power &25 dBm\\ \hline
    Transmitting link gain  &26 dB\\ \hline
    Transmitting power consumption  &1.8 W\\ \hline
    Receiving link gain  &33 dB\\ \hline
    Receiving power consumption &1 W \\ \hline
    Receiving noise factor &1.8 dB\\
    \hline \hline
\label{parameter0}
\end{tabular}
\end{table}

\begin{table}[!t]
\caption{The measured send attenuation and packet loss rate performance.}
\centering
\begin{tabular}{c c c}
	\hline Modulation mode, code rate& Send attenuation & Packet loss rate\\
	\hline \hline QPSK, 1/2 &106 dB &0.51\%\\
    \hline QPSK, 3/4 &106 dB&14.80\%\\
    \hline 16QAM, 1/2 &96 dB&0\\
    \hline 16QAM, 3/4 &96 dB&0.88\%\\
    \hline 64QAM, 2/3  &86 dB&0\\
    \hline 64QAM, 3/4  &86 dB&0.32\% \\
    \hline \hline
\label{parameter2}
\end{tabular}
\end{table}

Besides, the baseband card performs the self-loopback test over the coaxial line by adding an RF attenuator to the link and setting the AD9361's programmed digital attenuator, and the sum of the attenuations of the above two attenuators is defined as the send attenuation.
The measured send attenuation and packet loss rate under different modulation modes and code rates are shown in Tab. \ref{parameter2}.
Note that the packet loss rate is the proportion of frames that fail to pass CRC checksum at the receiver.
The terminal's adjacent channel leakage power ratio (ACLR) of the upper adjacent channel and the lower adjacent channel are separately equal to 48.19 dBc and 46.27 dBc, which meet the requirement of 3GPP TS36.141 standard.
\subsection{Static and Dynamic Power Consumption Analysis}
In this subsection, we test the static power consumption and the dynamic power consumption of the RIS, respectively.

{\bf Case I:}
We test the RIS's static power consumption $P_s$ versus the proportion of the conductive PIN Diode $r$ and then compute the power consumption of a single conductive PIN Diode.

{\bf Case II:}
We perform RIS codebook switching and test the whole RIS array's dynamic power consumption $P_d$ versus its codebook switching frequency $f$.

\begin{table}[!t]
\caption{The measured static power consumption $P_s$ versus the conduction ratio $r$.}
\centering
\label{static data}
\begin{tabular}{|c|c|c|c|c|c|c|c|c|c|}
	\hline $r$ &0&0.125&0.25&0.375&0.5&0.625&0.75&0.875&1 \\
	\hline $P_s$[W] &1.9836&2.0402&2.0981&2.1519&2.2119&2.271&2.3284&2.392&2.4536\\
	\hline
\end{tabular}
\end{table}

\begin{table}[!t]
\caption{The measured dynamic power consumption $P_d$ versus the codebook switching frequency $f$.}
\centering
\label{dynamic data}
\begin{tabular}{|c|c|c|c|c|c|c|c|c|c|c|c|c|c|}
	\hline $f$[Hz] &10K&20K&40K&60K&80K&100K&150K&200K&300K&400K&500K&600K&700K \\
	\hline $P_d$[W] &2.22&2.23&2.254&2.278&2.3&2.322&2.378&2.436&2.5487&2.6543&2.757&2.857&2.945\\
	\hline
\end{tabular}
\end{table}

The measured static power consumption $P_s$ of the RIS versus the proportion of the conductive PIN Diode $r$ is shown in Tab. \ref{static data}.
Note that if $r=0$, then all the PIN Diodes are cut-off and the corresponding $P_s$ is induced by the RIS controller, which is equal to 1.9836.
Through least square (LS) method, the quadratic relationship between $P_s$ and $r$ can be fitted as
\begin{align}
         P_s= 1.9386+0.0046r-4.3204*10^{-6}r^2.
\end{align}
We can see that the coefficient of the quadratic term is negligible, thus $P_s$ increases linearly with the increase of $r$.
The power consumption $P_{pin}$ induced by a single conductive PIN Diode can be computed as follows.
First, it should be pointed out that each PIN Diode is equipped with a corresponding current-limiting resistance, which will also result in power consumption.
The power consumption $P_{pin, res}$ of a conductive PIN Diode together with its corresponding current-limiting resistance can be computed as $P_{pin, res}= (2.4536-1.9836)/512 = 0.918$mW, where the denominator 512 is due to the fact that the RIS adopts two-bit phase quantization with each unit equipped with two PIN Diodes.
The power consumption $P_{res}$ of the current-limiting resistance is equal to $P_{res}= V_{res}^2/R=0.32$mW, where $V_{res}=0.4$V is the voltage over the resistor and $R=400\Omega$ is the resistance value of the resistor.
Thus, $P_{pin}$ can be computed as $P_{pin}= P_{pin, res}-P_{res}=0.598$mW.

The measured dynamic power consumption $P_d$ of the whole RIS array versus the RIS codebook switching frequency $f$ is shown in Tab. \ref{dynamic data}.
Through LS method, the relationship between $P_d$ and $f$ can be fitted as
\begin{align}
         P_d= 2.2141+0.001f+5.1469*10^{-7}f^2.
\end{align}
It can be seen that the $P_d$ increases linearly with the increase of $f$.
Specifically, as $f$ increases 1KHz, $P_d$ will increase 1mW, which is relative large.
Thus, if the RIS codebook is switching at a relative high frequency, the dynamic power consumption should also be taken into account when evaluating the total power consumption of the RIS.
\subsection{Beam Focusing Performance}
In this subsection, the following case is considered in order to compare the differences between the RIS and the metal plate in terms of  the beam focusing performance.

{\bf Case III:}
We test the received power at some specific points within an indoor environment, where the metal plate and the RIS are separately deployed and the RIS codebook is designed to reflect the incident signal towards the desired focus point.

\begin{figure}[!t]
	\centering
	\includegraphics[width=80mm]{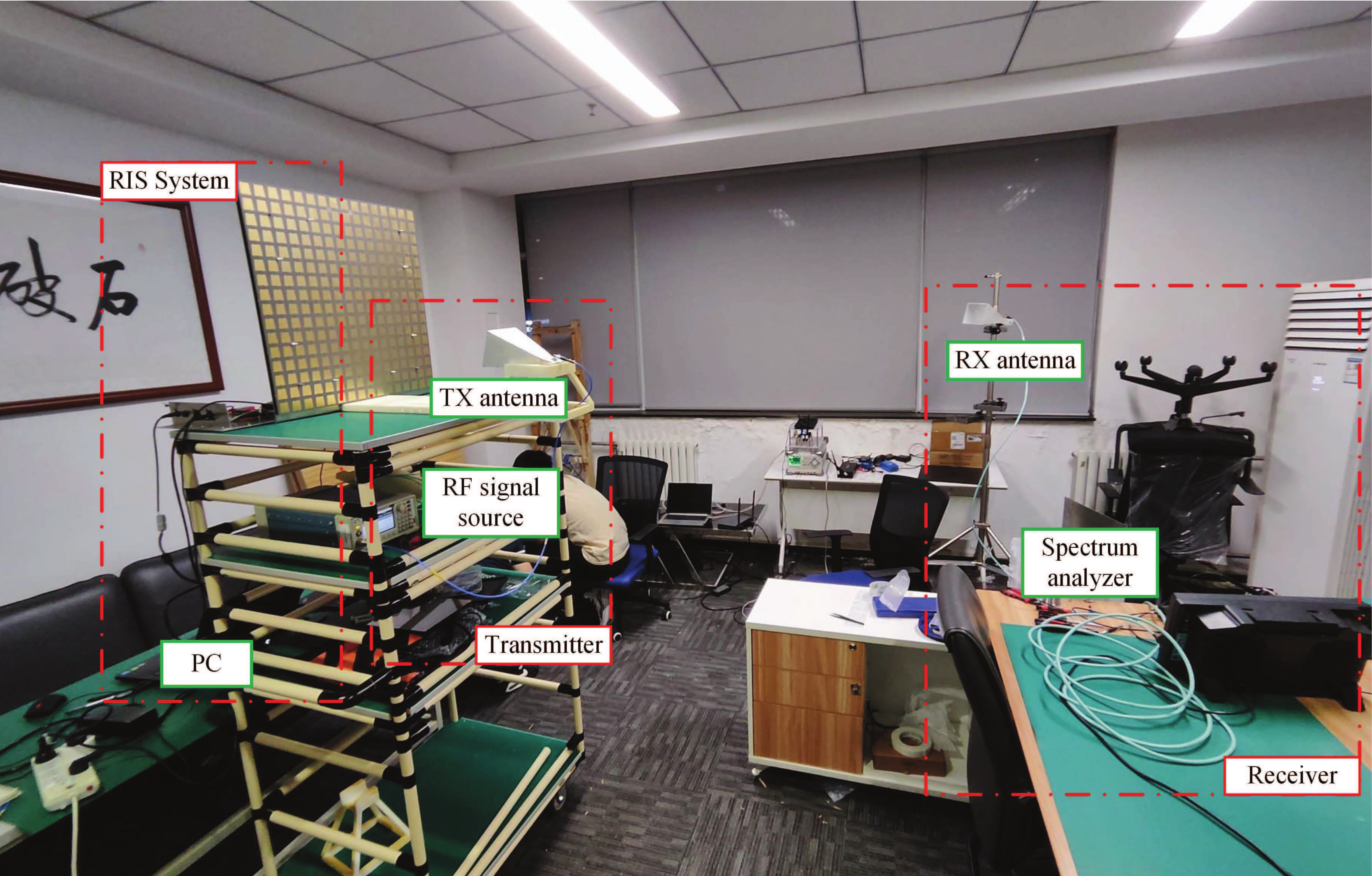}
	\caption{{Test environment in terms of the RIS's beam focusing performance.}}
	\label{beamfocusing}
\end{figure}

\begin{figure}[!t]
	\centering
	\includegraphics[width=90mm]{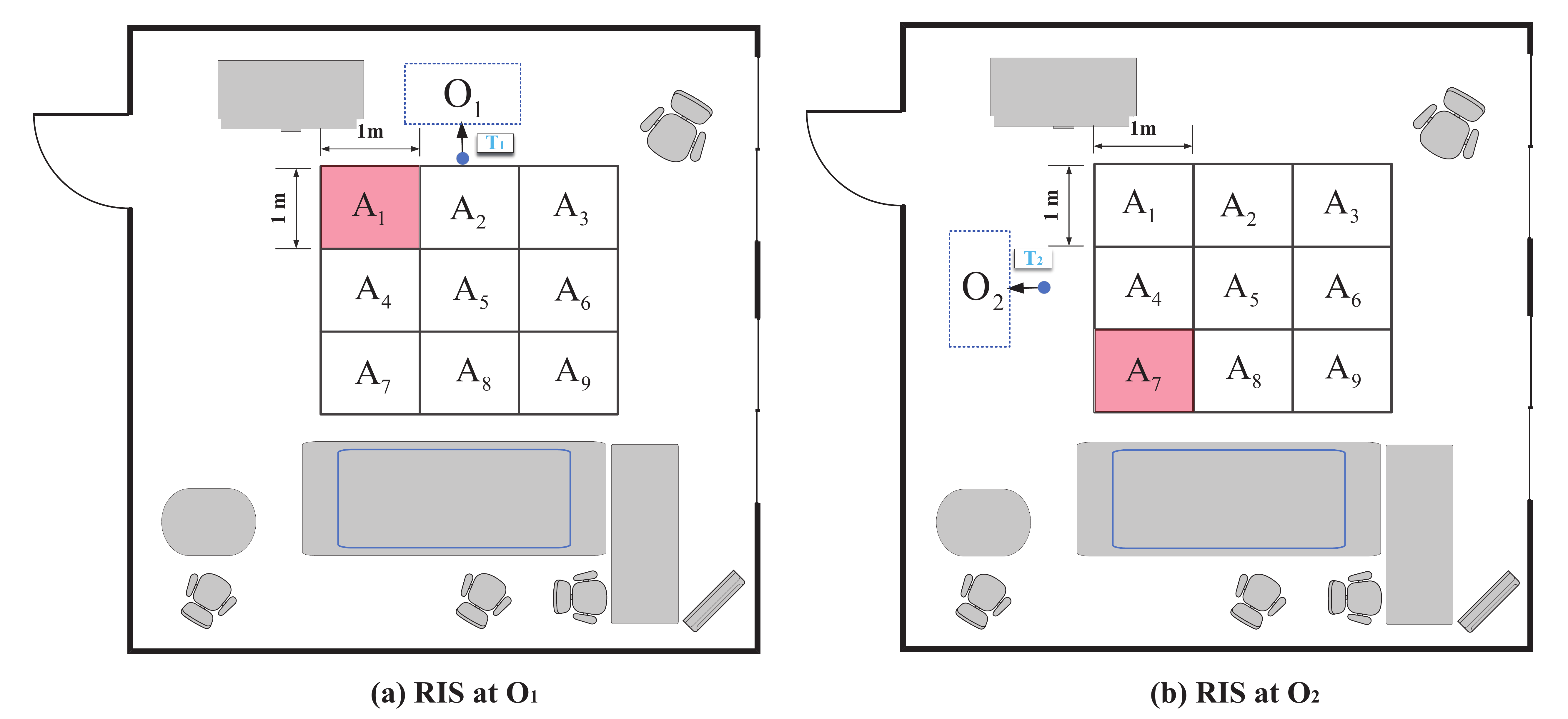}
	\caption{{Plane topology in terms of the RIS's beam focusing performance.}}
	\label{beamfocusing_topology}
\end{figure}

The test environment and the plane topology are shown in Fig. \ref{beamfocusing} and Fig. \ref{beamfocusing_topology}, respectively.
Specifically, we select a complex indoor environment and set nine test points, which are labeled as `$\text A_1$', `$\text A_2$', ..., `$\text A_9$', respectively.
Note that the locations of these test points are marked in Fig. \ref{beamfocusing_topology}.
Moreover, two locations labeled as `$\text O_1$' and `$\text O_2$' are chosen for both the RIS and the metal plate.
Correspondingly, the transmitter is located at `$\text T_1$' and `$\text T_2$', which are 0.5m away from `$\text O_1$' and `$\text O_2$', respectively.
When the RIS is deployed, we design the RIS codebook according to \eqref{singlebeam} to reflect the incident signal towards `$\text A_1$' and `$\text A_7$', respectively.
We test the received power at `$\text A_1$'--`$\text A_9$' by moving the receiver to these tests points in sequence.
Besides, we set $P_t=-11.5$ dBm in this case.

The corresponding test results are displayed in Tab. \ref{dynamic data1}.
We can see that when the metal plate is deployed, the received power at `$\text A_1$'--`$\text A_9$' is broadly consistent.
Specifically, the maximum difference of the received power among these test points is about 3 dBm, which is relative small.
By comparison, the deployment of the RIS results in a significant enhancement in terms of the received power at the desired focus point `$\text A_1$' or `$\text A_7$'.
Specifically, the received power at the desired focus point is about 20 dBm higher than that at the other test points.
Thus, with respect to the beam focusing performance, the RIS significantly outperforms the metal plate.

\begin{table}[!t]
\caption{The measured $P_r$ at `$\text A_1$'--`$\text A_9$' under different conditions.}
\centering
\resizebox{\textwidth}{!}{
\label{dynamic data1}
\begin{tabular}{|c|c|c|c|c|c|c|c|c|c|}
	\hline \diagbox{Conditions}{Test points}&`$\text A_1$'&`$\text A_2$'&`$\text A_3$'&`$\text A_4$'&`$\text A_5$'&`$\text A_6$'&`$\text A_7$'&`$\text A_8$'&`$\text A_9$' \\
	\hline metal plate at `$\text O_1$'&-44 dBm&-44.5 dBm&-45.3 dBm&-45.3 dBm&-46 dBm&-42 dBm&-44 dBm&-44.5 dBm&-46 dBm \\
    \hline RIS at `$\text O_1$'&\textcolor{red}{-32.8 dBm}&-50 dBm&-44.2 dBm&-40.8 dBm&-47.8 dBm&-50.2 dBm&-50.6 dBm&-50.1 dBm&-51 dBm \\
	\hline metal plate at `$\text O_2$'&-45.2 dBm&-44 dBm&-46.5 dBm&-45.7 dBm&-45.3 dBm&-45.7 dBm&-44 dBm&-43.7 dBm&-44.3 dBm \\
    \hline RIS at `$\text O_2$'&-44.7 dBm&-50.3 dBm&-53 dBm&-47.2 dBm&-47.9 dBm&-50.3 dBm&\textcolor{red}{-32.9 dBm}&-38.6 dBm&-53 dBm \\
	\hline
\end{tabular}}
\end{table}
\subsection{Channel Stationarity}
In this subsection, the following cases are tested in order to evaluate the effect of the RIS on the channel stationarity.

\begin{figure}[!t]
	\centering
	\subfigbottomskip=2pt
	\subfigcapskip=-5pt
	\subfigure[]{
		\includegraphics[width=0.41\linewidth]{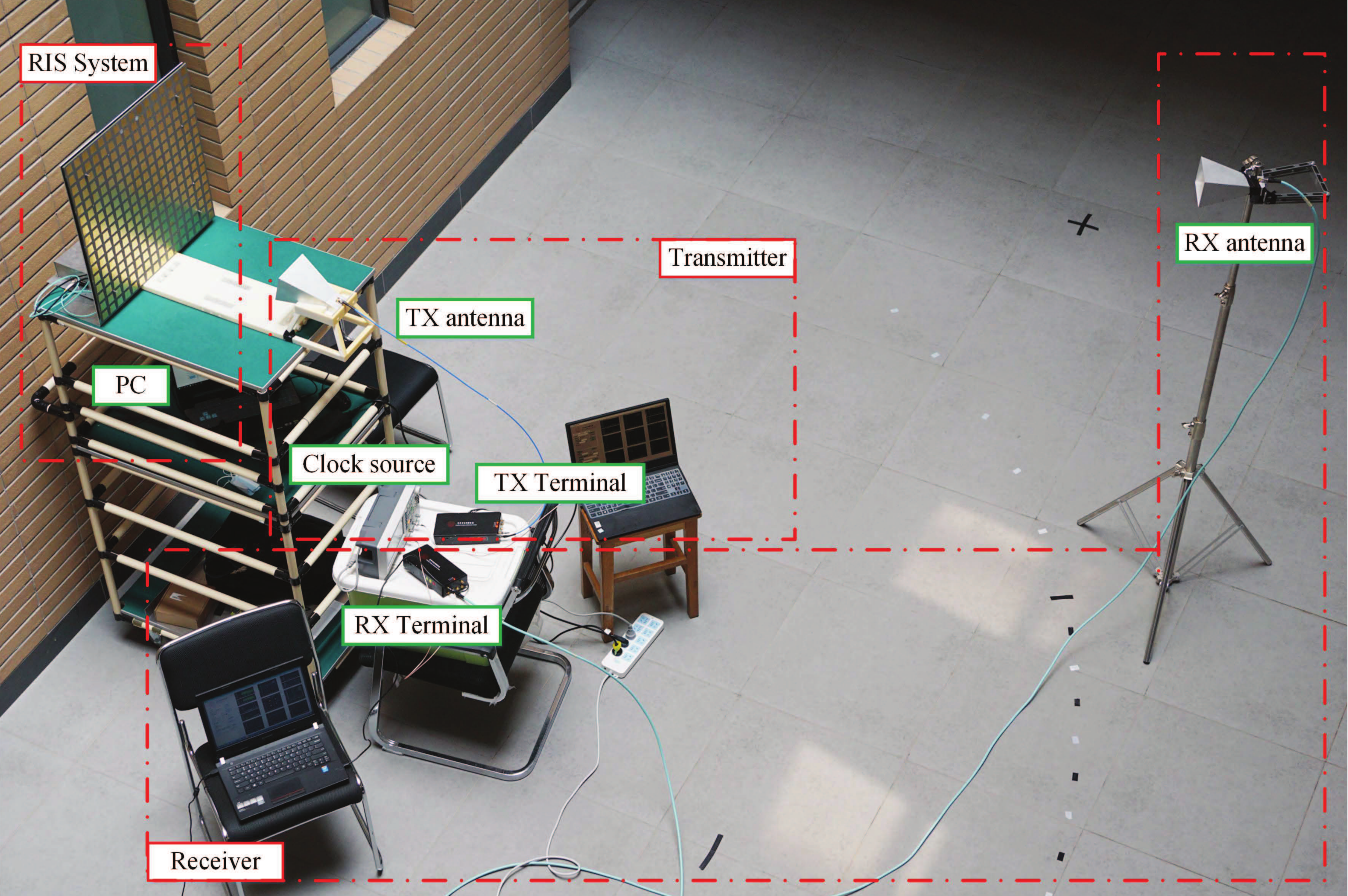}}
	\subfigure[]{
		\includegraphics[width=0.56\linewidth]{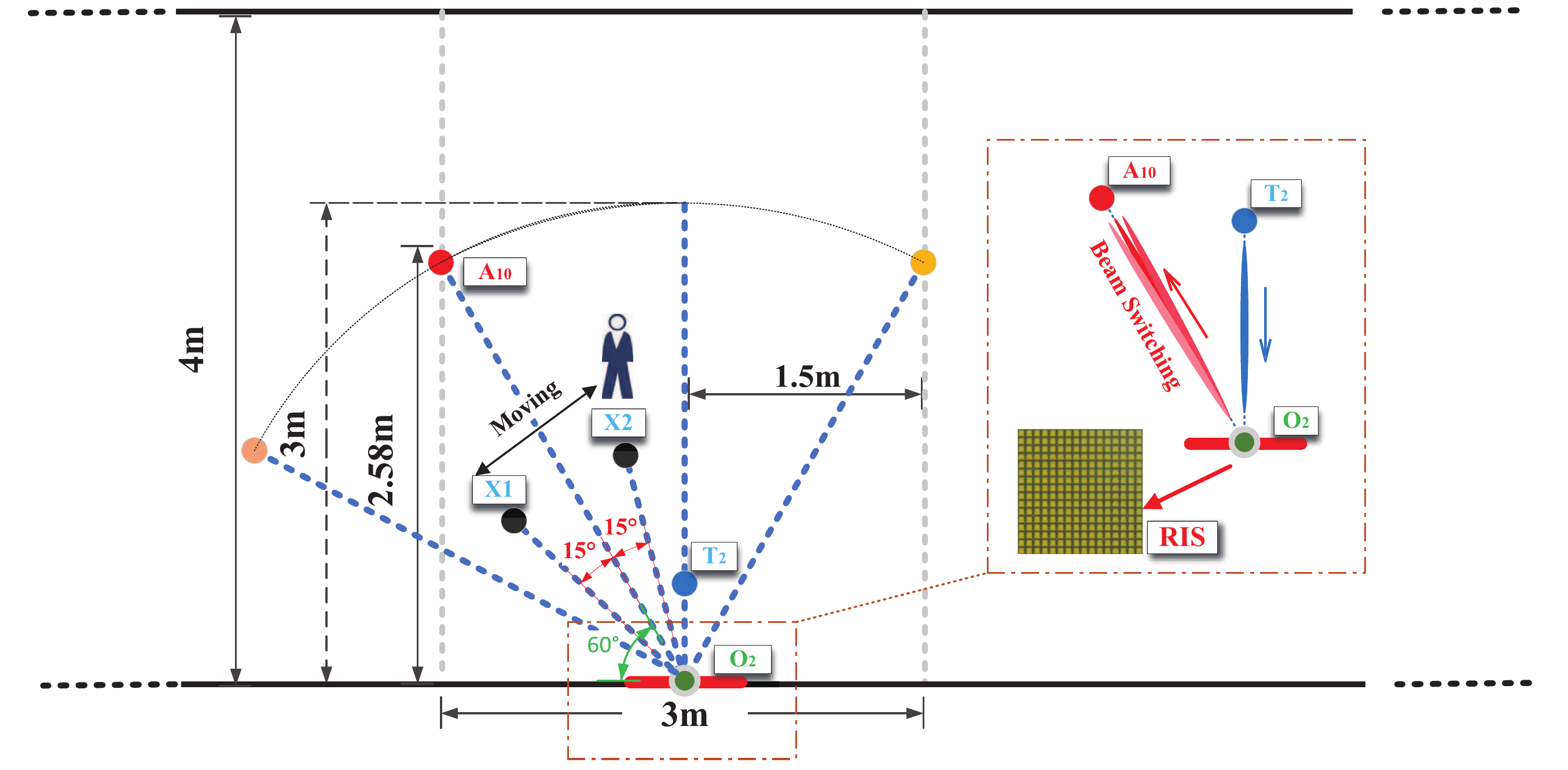}}
\caption{Channel stationary measurement. (a) Test environment. (b) Plane topology.}
\label{stationary}
\end{figure}

{\bf Case IV:}
We test the normalization difference of equivalent channel's statistic characteristics, where the metal plate and the RIS are separately deployed and the RIS is designed to reflect the incident signal towards the receiver.

{\bf Case V:}
We test the normalization difference of equivalent channel's statistic characteristics, where the RIS is deployed.
Moreover, the RIS is performing beam switching and its reflected beam will scan an angle range centered on the receiver.

{\bf Case VI:}
We test the normalization difference of equivalent channel's statistic characteristics, where the metal plate and the RIS are separately deployed and the RIS is designed to reflect the incident signal towards the receiver.
Moreover, an obstacle is moving around the receiver, which induces varing environment.

\begin{figure}[!t]
	\centering
	\subfigbottomskip=2pt
	\subfigcapskip=-5pt
	\subfigure[]{
		\includegraphics[width=0.45\linewidth]{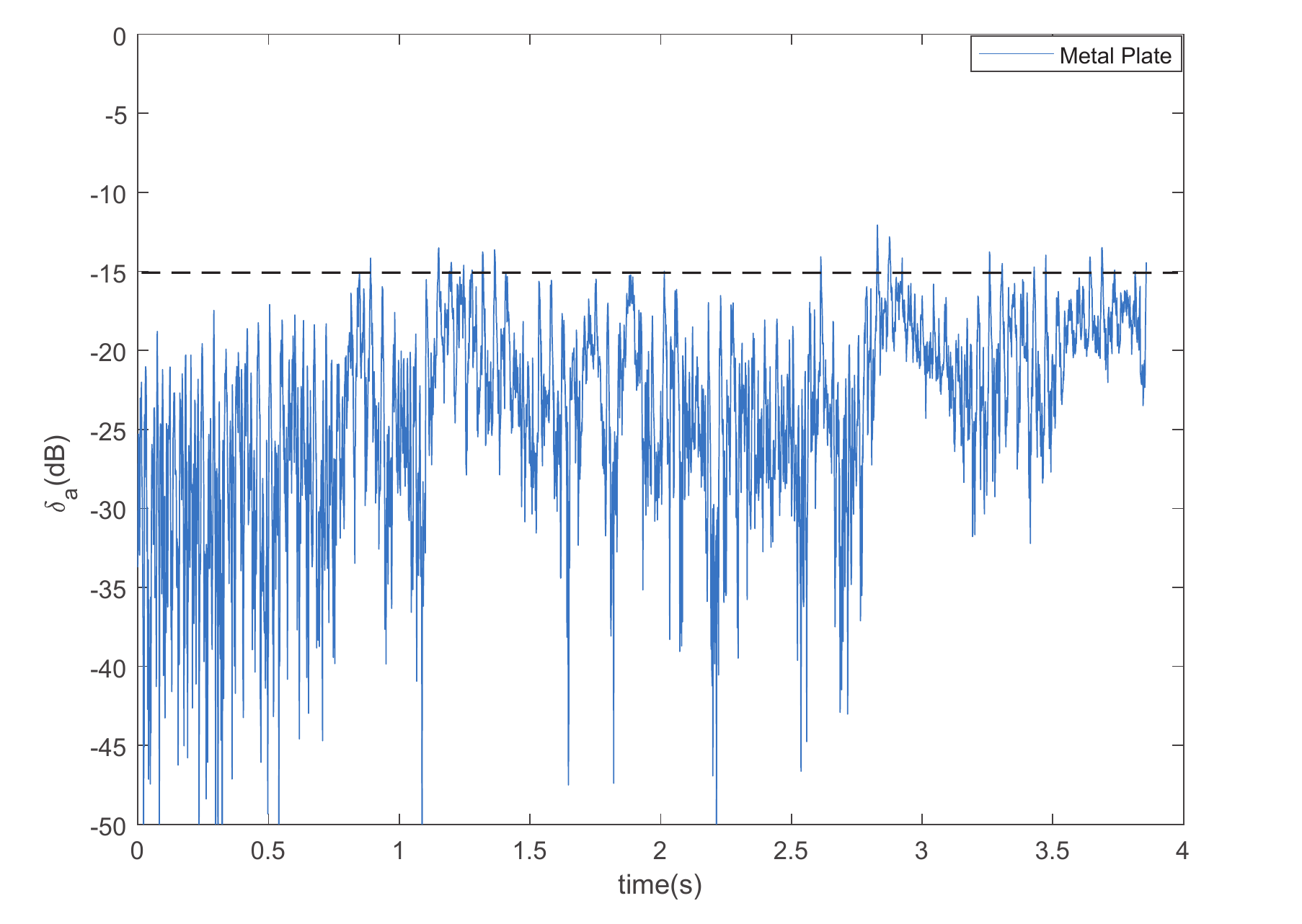}}
	\subfigure[]{
		\includegraphics[width=0.45\linewidth]{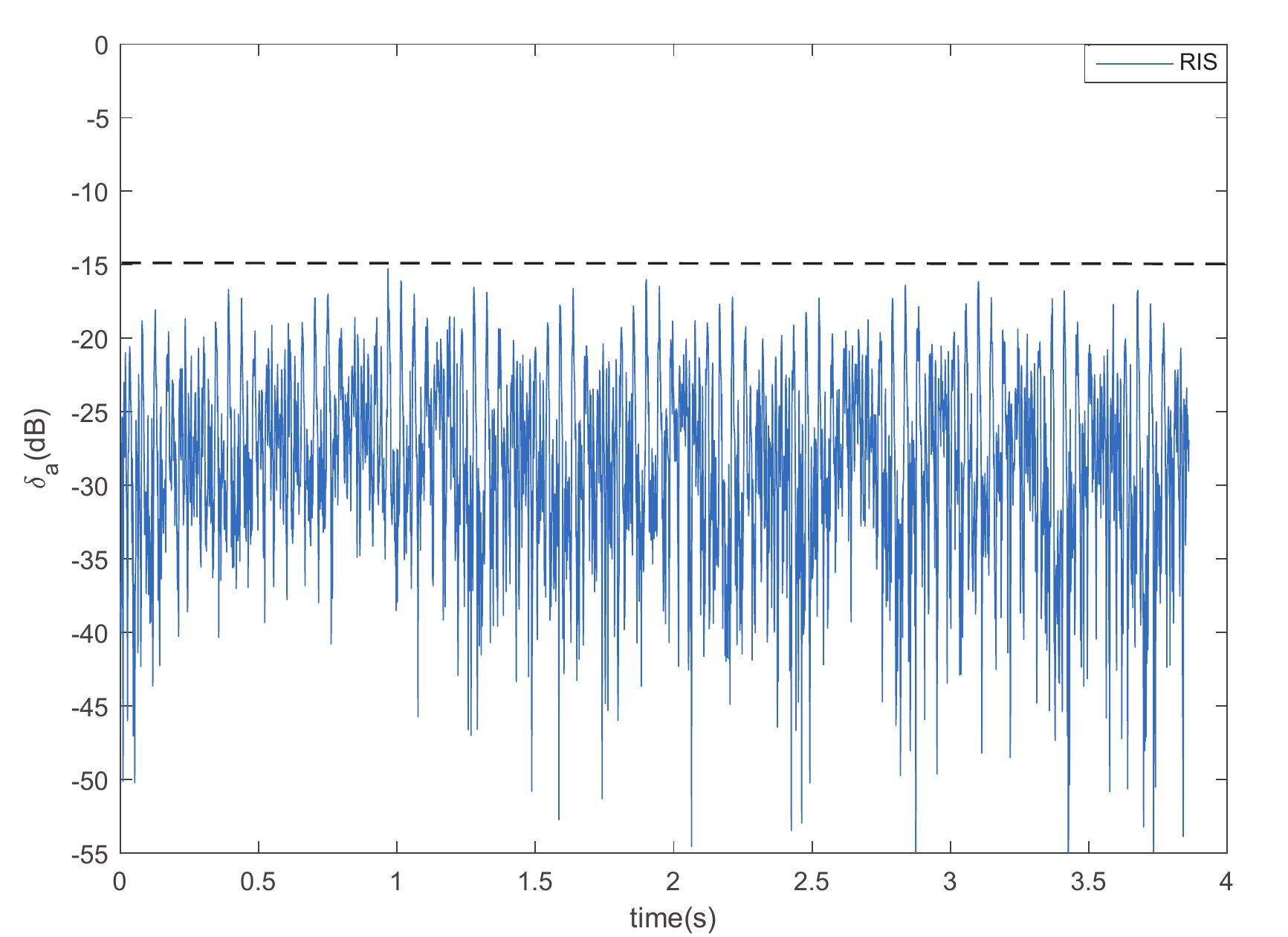}}
	\caption{The variation of $\delta_a[t]$ with time $t$. (a) The metal plate is deployed. (b) The RIS is deployed.}
\label{Stationarity_1}
\end{figure}

The test environment and the plane topology are separately illustrated in Fig. \ref{stationary}(a) and Fig. \ref{stationary}(b).
It can be seen that the RIS and the metal plate are deployed at point `$\text O_2$', while the transmitter and the receiver are separately positioned at `$\text T_2$' and `$\text A_{10}$'.
Moreover, we use a clock source RIGOL-DG4202 in this case to ensure the accuracy of the channel measurement, which is coordinated with the BICT-40N terminal to evaluate the channel stationary.
We first consider the case where the RIS codebook is designed to reflect the incident signal towards `$\text A_{10}$' by setting $\theta_r=30^{\circ}$ and $\phi_r=0^{\circ}$ in \eqref{singlebeam}.
We take $a[t]$ as an example to evaluate $\delta_x[t]$ in \eqref{delta}, which is denoted as $\delta_a[t]$ here.
The test results are shown in Fig. \ref{Stationarity_1}, which shows the variation of $\delta_a[t]$ with time $t$.
The threshold is set as -15 dB.
We can see that if the metal plate is deployed, then the channel is non-stationary since $\delta_a[t]$ will exceed -15 dB at certain times.
However, if the RIS is deployed, then $\delta_a[t]$ in the entire time duration is below -15 dB, i.e., the channel becomes stationary with the aid of the RIS.

\begin{figure}[!t]
	\centering
	\subfigbottomskip=2pt
	\subfigcapskip=-5pt
	\subfigure[]{
		\includegraphics[width=0.45\linewidth]{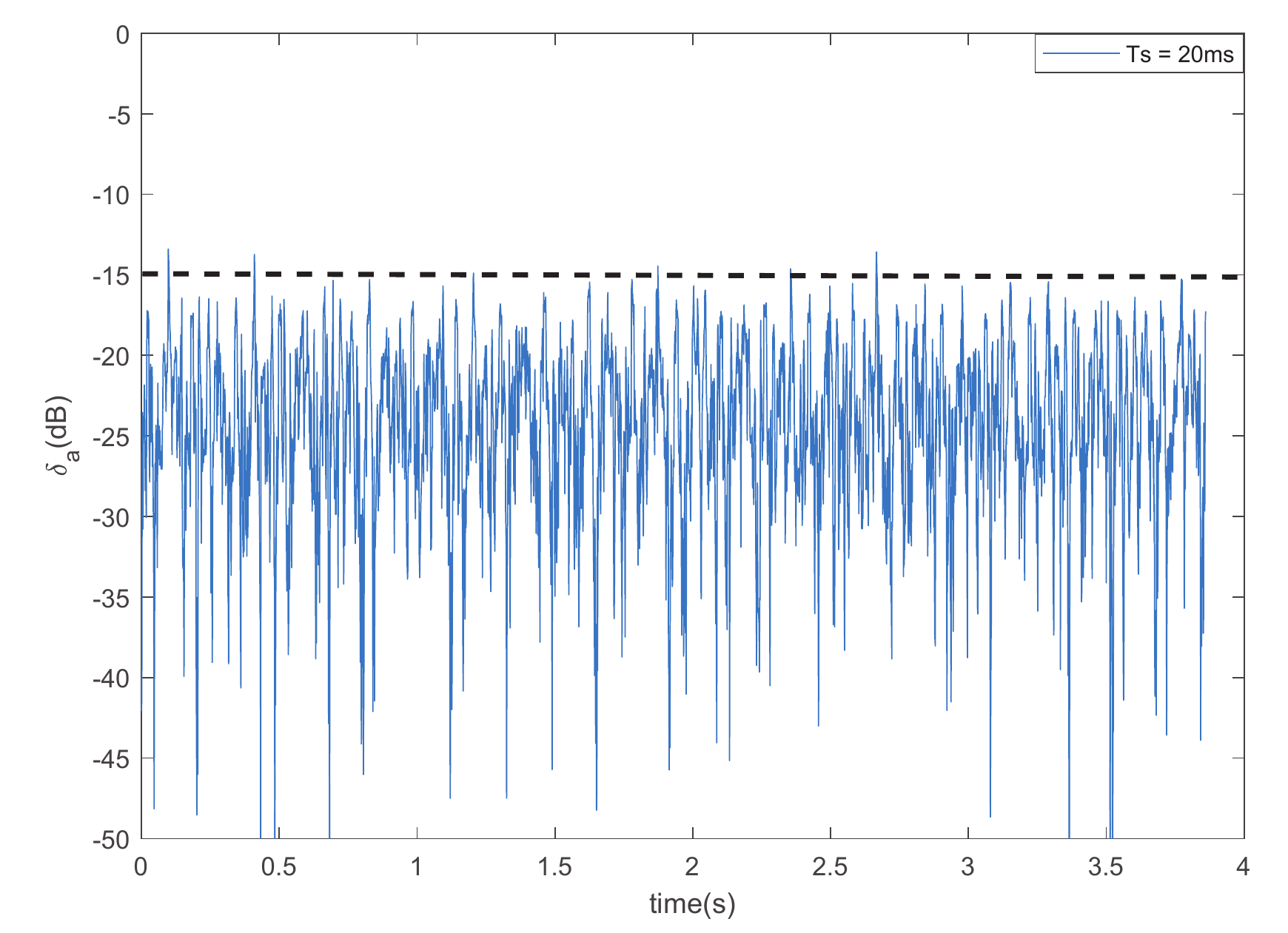}}
	\subfigure[]{
		\includegraphics[width=0.45\linewidth]{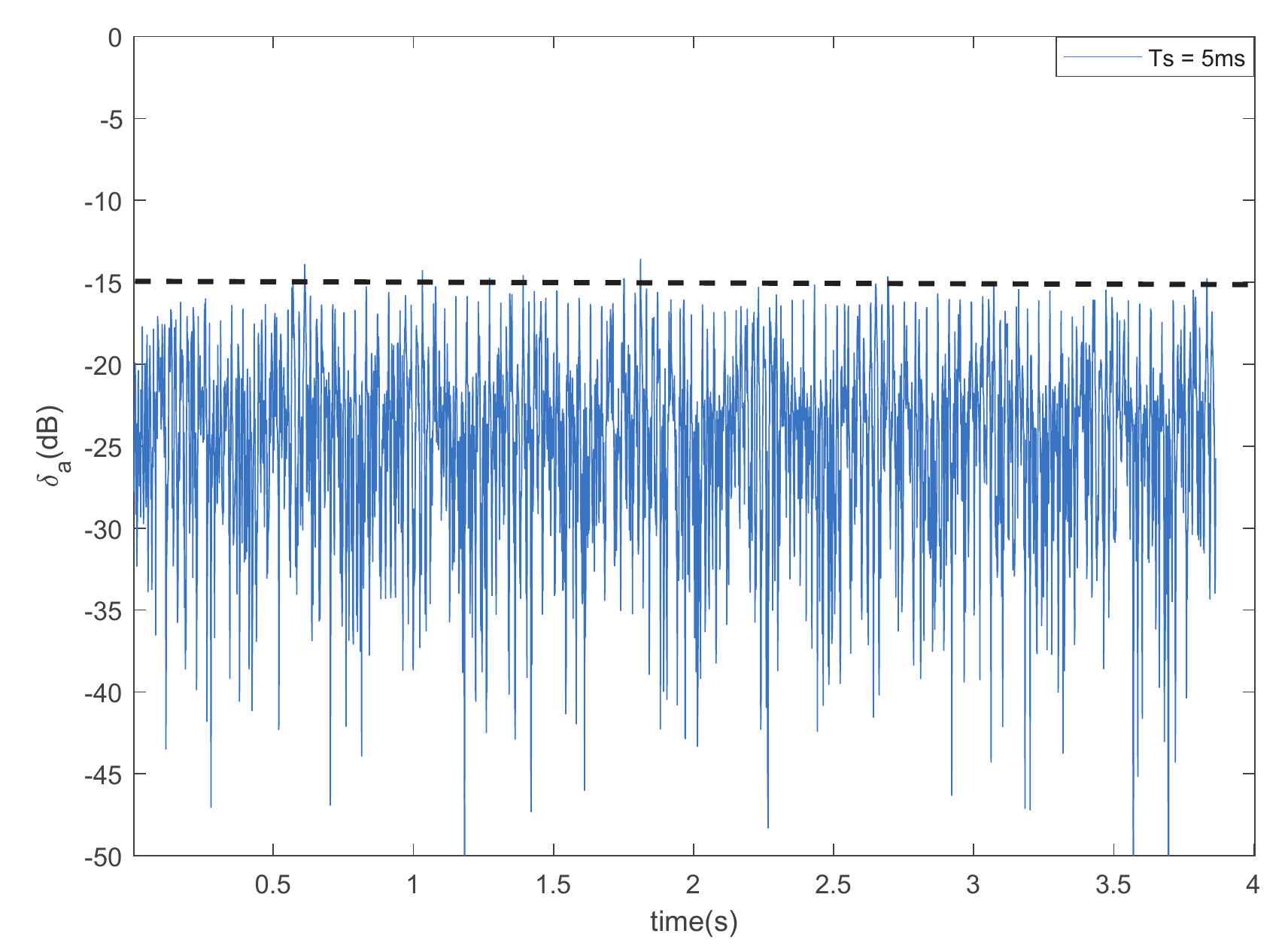}}\\
	\subfigure[]{
		\includegraphics[width=0.45\linewidth]{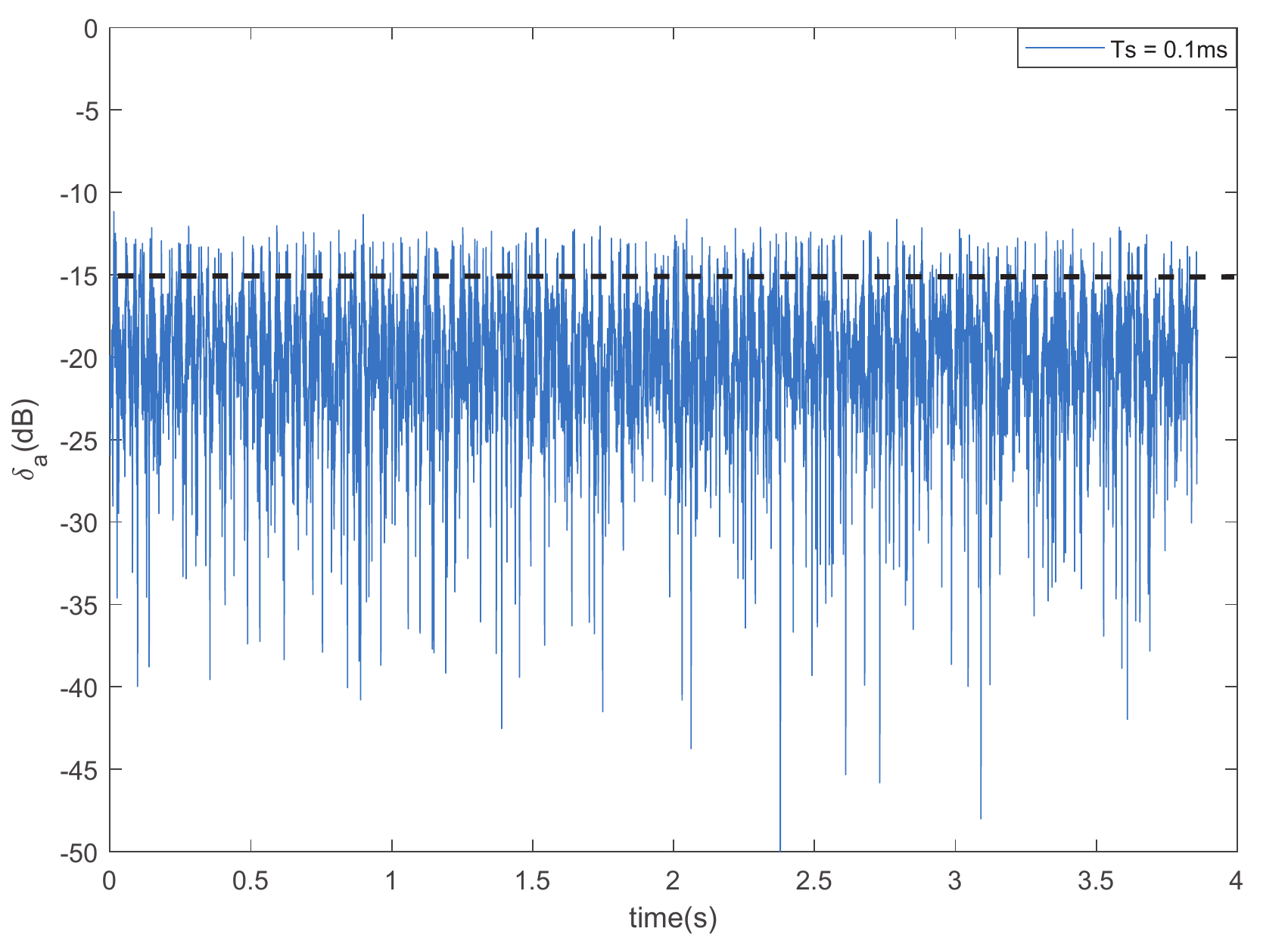}}
	\subfigure[]{
		\includegraphics[width=0.45\linewidth]{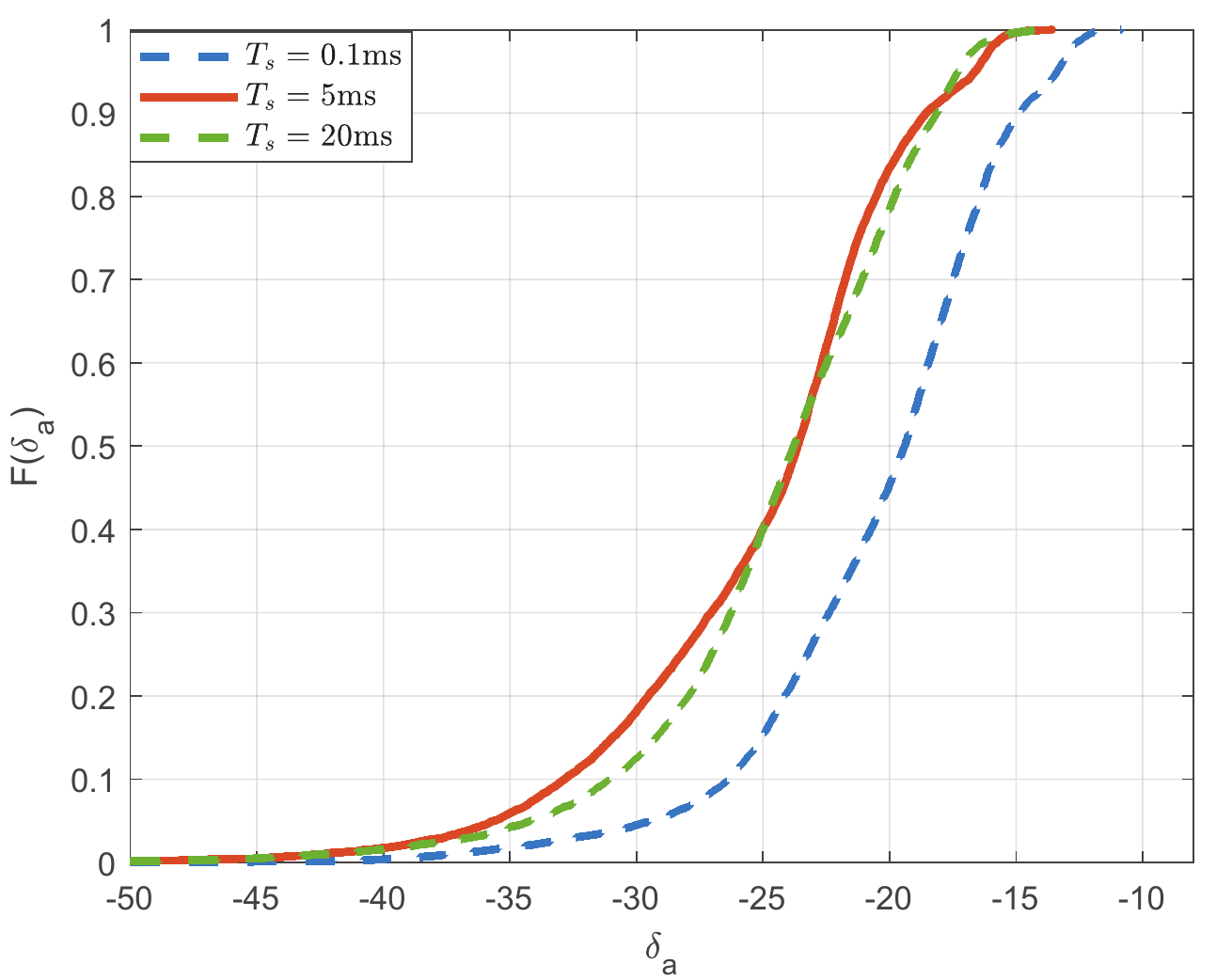}}
	\caption{The variation of $\delta_{a}[t]$ with time $t$ under different $T_{\text s}$. (a) $T_{\text s}$ = 20ms. (b) $T_{\text s}$ = 5ms. (c) $T_{\text s}$ = 0.1ms. (d) CDF curves.}
\label{Stationarity_2}
\end{figure}
\begin{figure}[t]
	\centering  
	\subfigbottomskip=2pt 
	\subfigcapskip=-5pt 
	\subfigure[]{
		\includegraphics[width=0.42\linewidth]{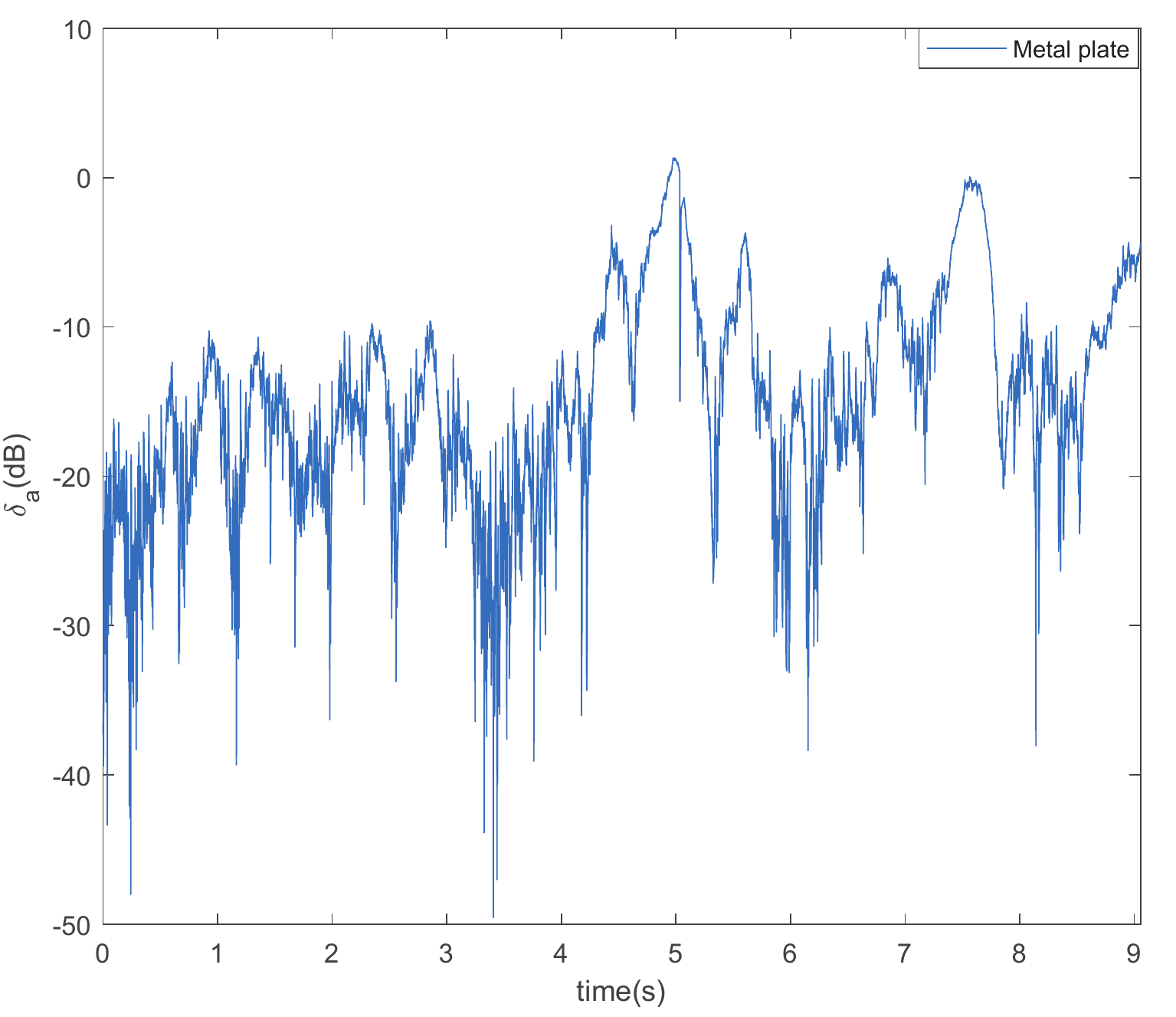}}
	\subfigure[]{
		\includegraphics[width=0.43\linewidth]{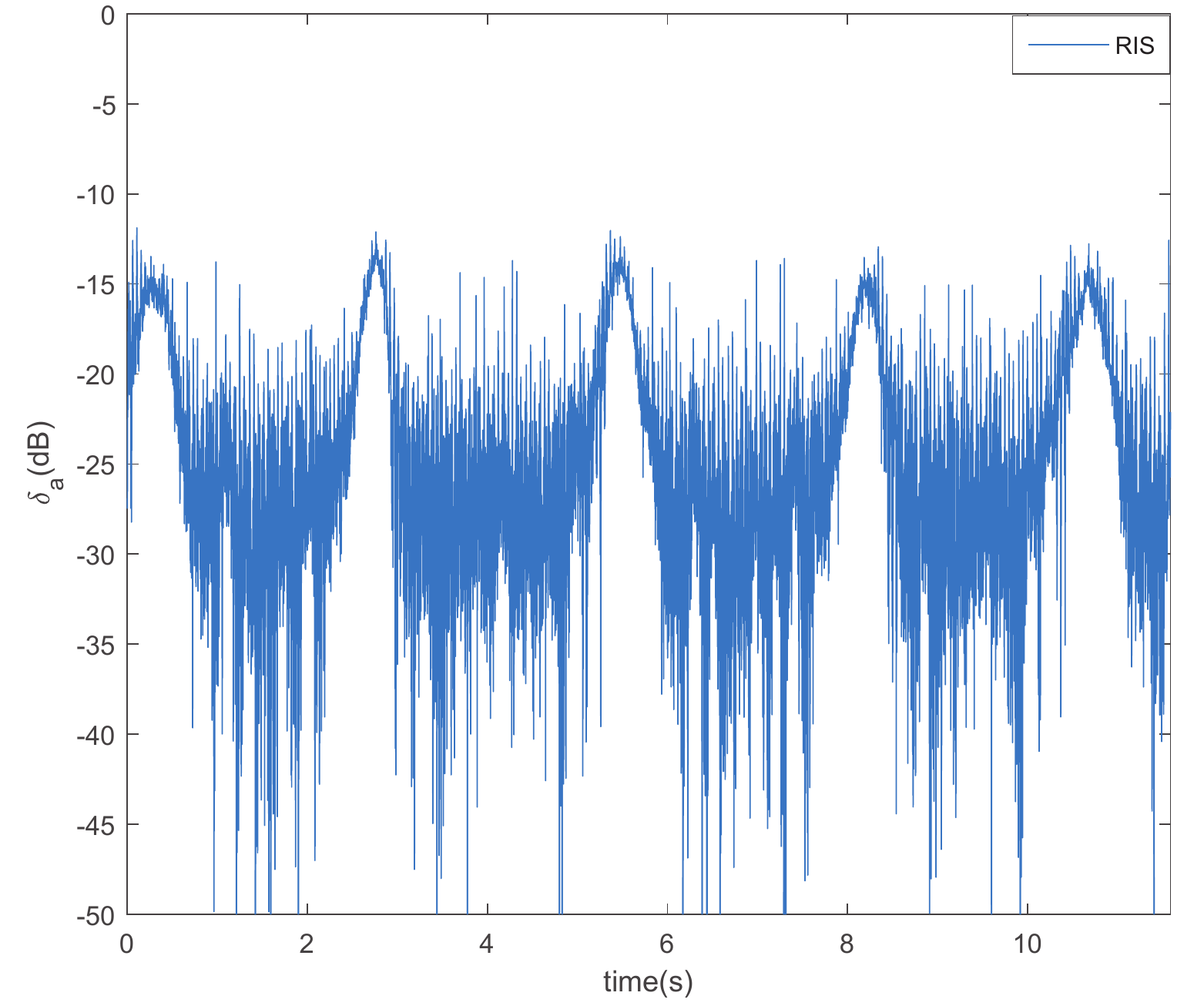}}
	\caption{The variation of $\delta_a[t]$ with time $t$ under varying environment. (a) The metal plate is deployed. (b) The RIS is deployed.}
\label{Stationarity_3_Mobile scene}
\end{figure}

To validate the effect of the RIS beam switching on the channel stationary, we assign nine different codewords to the RIS.
Specifically, these nine codewords are separately designed according to \eqref{singlebeam}, where $\theta_r \in \{26^{\circ}, 27^{\circ}, 28^{\circ}, 29^{\circ}, 30^{\circ}, 31^{\circ}, 32^{\circ}, 33^{\circ}, 34^{\circ}\}$.
As a result, during the RIS beam switching, its reflected beam periodically scan the angle range $[26^{\circ},34^{\circ}]$ centered on `$\text A_{10}$'.
Here, the beam switching period is denoted as $T_{\text s}$, and we choose three values of $T_{\text s}$, including 20ms, 5ms and 0.1ms.
Fig. \ref{Stationarity_2} shows the variation of $\delta_a[t]$ with time $t$ under different $T_s$ and the corresponding cumulative distribution function (CDF) curves.

It can be seen from Fig. \ref{Stationarity_2} that when $T_s$ is set as 20ms or 5ms, almost all $\delta_a[t]$ within the time duration are less than -15 dB and thus the RIS channel is stationary.
Thus, a low beam switching frequency has little impact on the stationarity of the RIS channel.
However, as $T_{\text s}$ continues to decrease, the RIS channel will become non-stationary.
Specifically, when $T_{\text s}$ is 0.1ms, about 10$\%$ $\delta_a[t]$ within the time duration are greater than $-15$ dB.
This indicates that if the RIS beam switching frequency is too high, then the stationarity of the RIS channel will be destroyed.

In addition, Fig. \ref{Stationarity_3_Mobile scene} illustrates the impact of the RIS on the channel stationary under varying environment.
In this case, an obstacle is moving at a speed of 1 m/s between point `$\text X_1$' and point `$\text X_2$' in Fig. \ref{stationary}(b), which simulates the variation of the actual environment.
The RIS codebook is designed to reflect the incident signal towards `$\text A_{10}$'.
For one thing, it can be seen that the environmental variation will seriously destroy the channel stationary.
For another, compared to the scene where the metal plate is deployed, the channel stationary is improved to some extent with the aid of the RIS.

\subsection{Imaging Interference}
\begin{figure}[!t]
	\centering
	\subfigbottomskip=2pt
	\subfigcapskip=-5pt
	\subfigure[]{
		\includegraphics[width=0.41\linewidth]{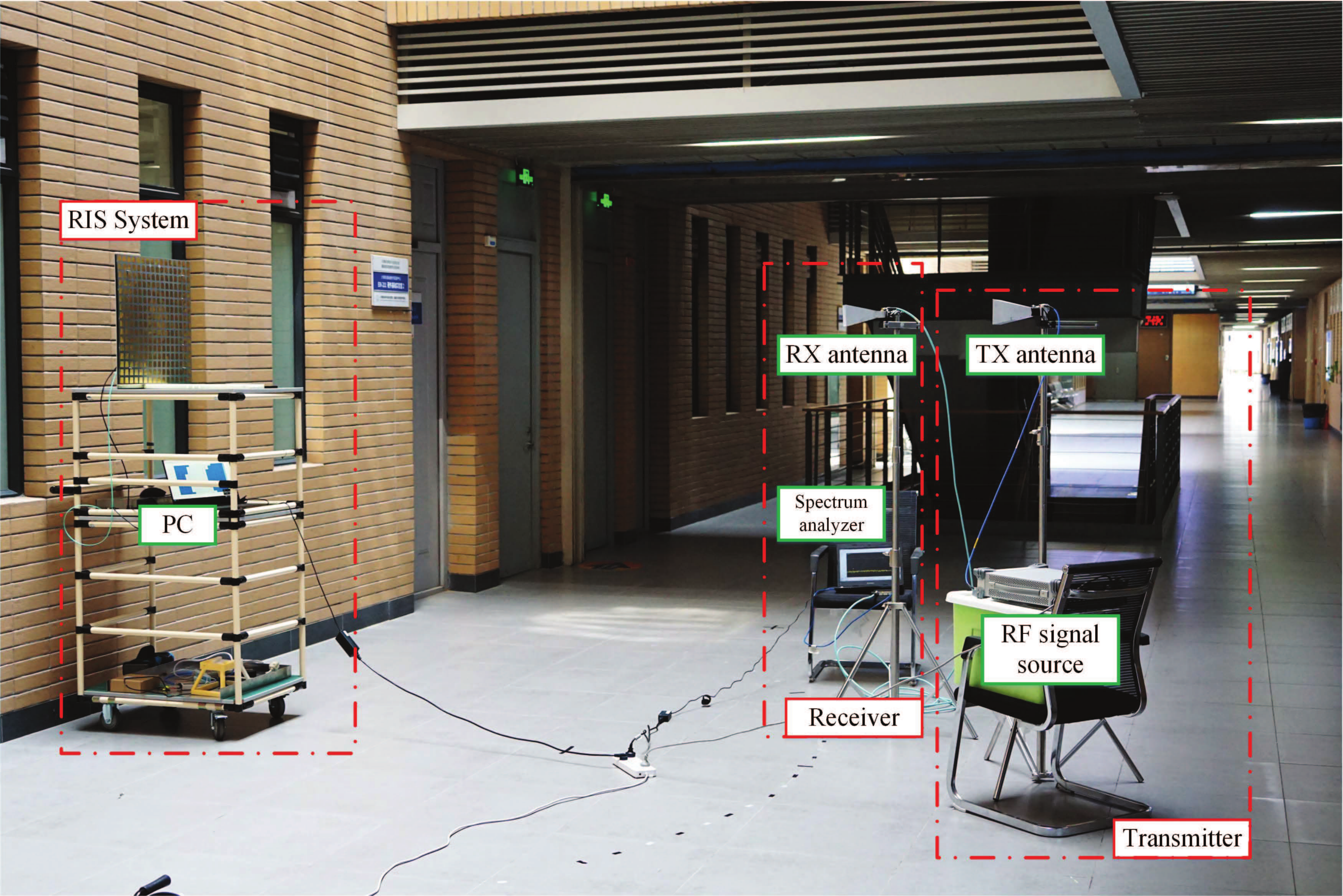}}
	\subfigure[]{
		\includegraphics[width=0.56\linewidth]{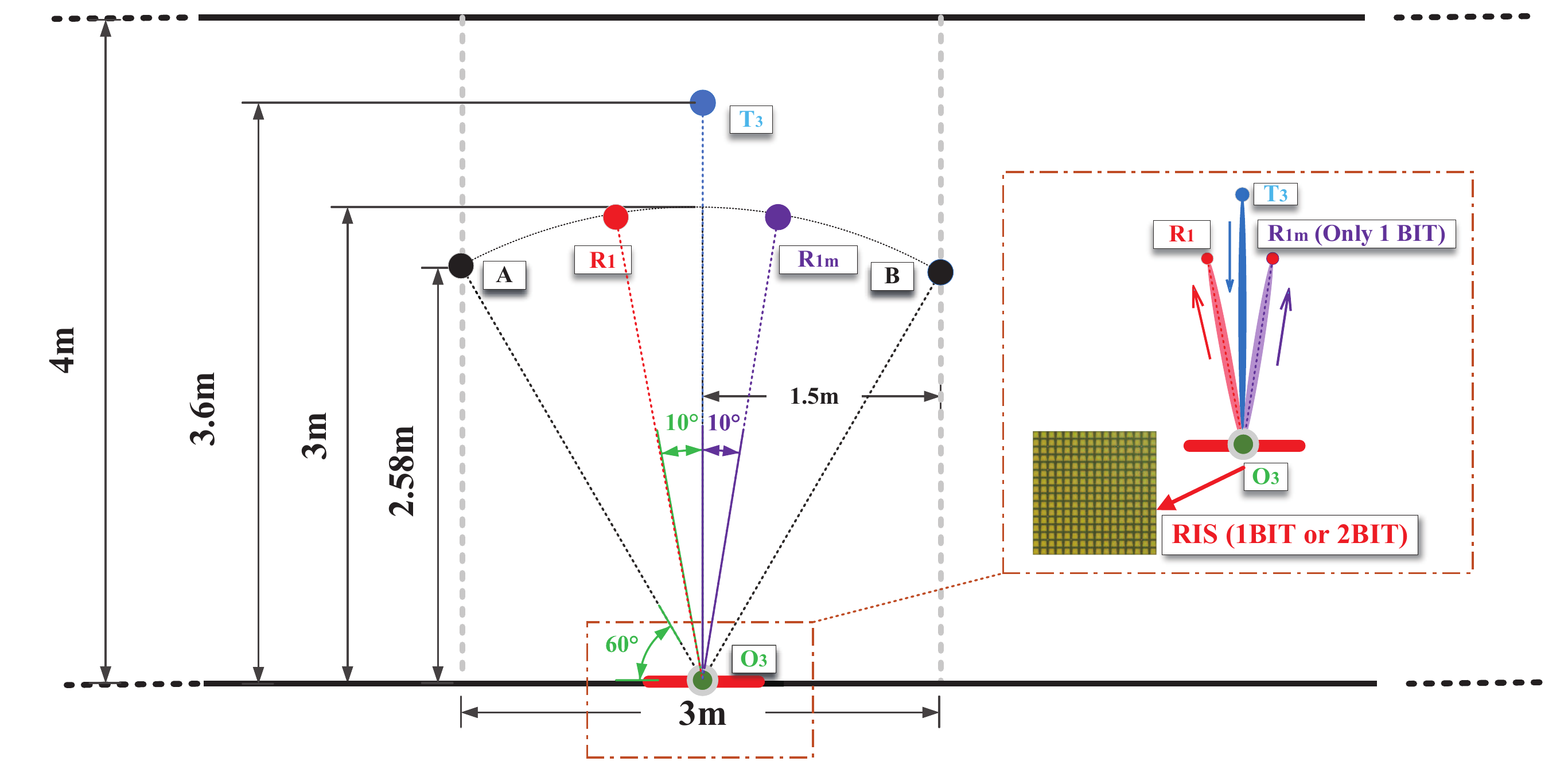}}
\caption{Imaging interference measurement. (a) Test environment. (b) Plane topology.}
\label{imaging_inte}
\end{figure}
In this subsection, the following case is considered in order to verify the existence and further test the main influencing factors of the imaging interference in RIS-aided systems.

{\bf Case VII:}
The RIS is deployed and is designed to reflect the incident signal towards the target receiver.
Then, we test the angle power spectrum of the reflected signal to observe the level of the imaging interference, where two different $d_t$ are selected and the quantization bit number of the RIS unit is separately set as one and two.

The test environment and the plane topology are separately shown in Fig. \ref{imaging_inte}(a) and Fig. \ref{imaging_inte}(b).
The transmitter and the RIS are located at point `$\text T_3$' and point `$\text O_3$', respectively.
We set $h_t=1.8$m in this case.
Moreover, we design the RIS codebook according to \eqref{singlebeam} in order to reflect the incident signal towards point `$\text R_1$', where $(\theta_r, \phi_r)$ is set as $(10^{\circ},0^{\circ})$.
Note that the mirror image of `$\text R_1$' is located at point `$\text R_{1m}$', whose exit angle is $(10^{\circ},180^{\circ})$.
Specifically, we use one-bit phase quantization and two-bit phase quantization, respectively.
Moreover, $d_t$ is separately set as 0.5m and 3.6m.
Then, we observe the angle power spectrum by moving the receiver along the dotted line in Fig. \ref{imaging_inte}(a), which corresponds to the arc $\overset{\frown} {\text{AB}}$ in Fig. \ref{imaging_inte}(b).

\begin{figure}[!t]
	\centering
	\includegraphics[width=100mm]{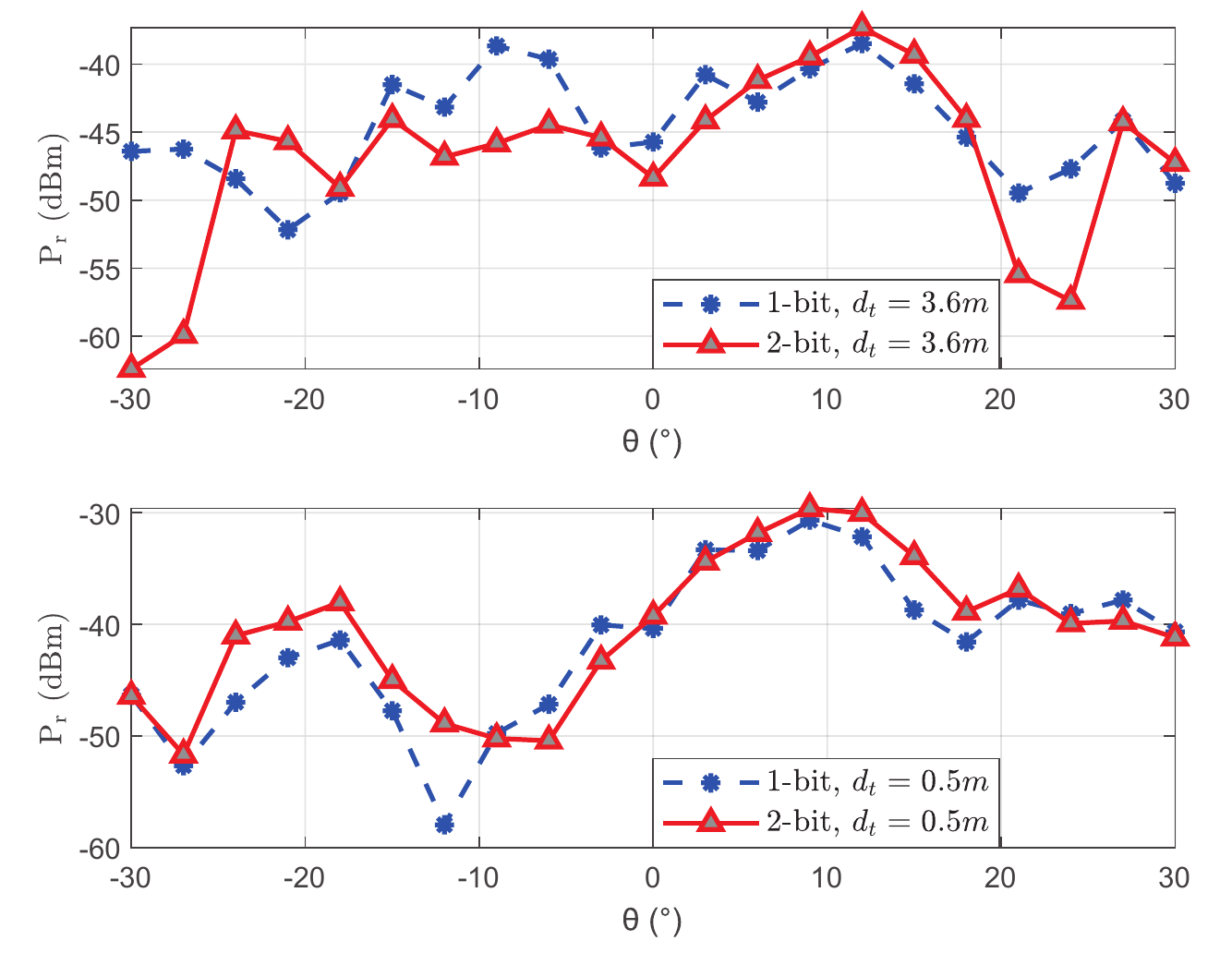}
	\caption{{The angular power spectrum in terms of the image interference measurement.}}
	\label{imaging_result}
\end{figure}
\begin{figure}[!t]
	\centering
	\subfigbottomskip=2pt
	\subfigcapskip=-5pt
	\subfigure[]{
		\includegraphics[width=0.35\linewidth]{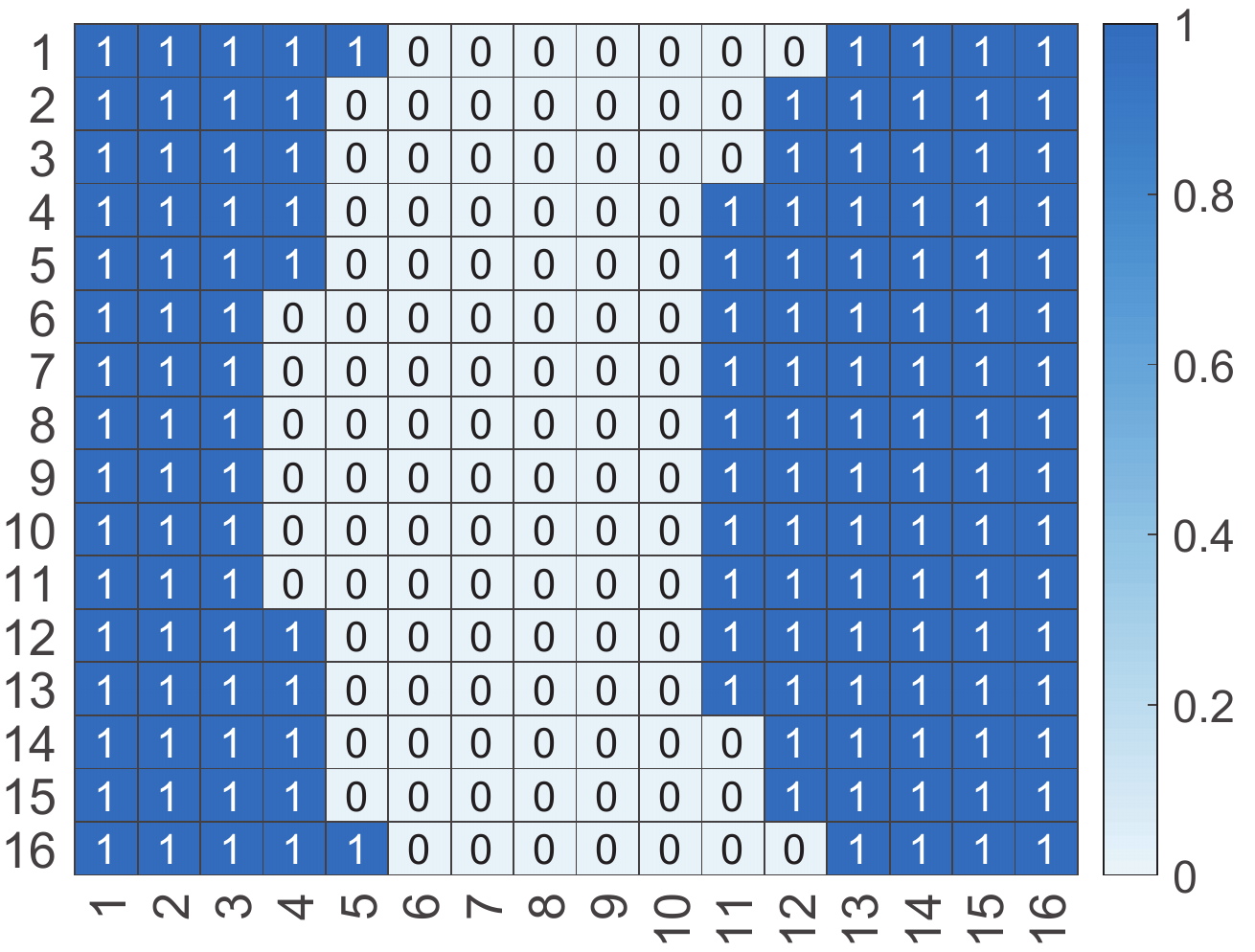}}
	\subfigure[]{
		\includegraphics[width=0.35\linewidth]{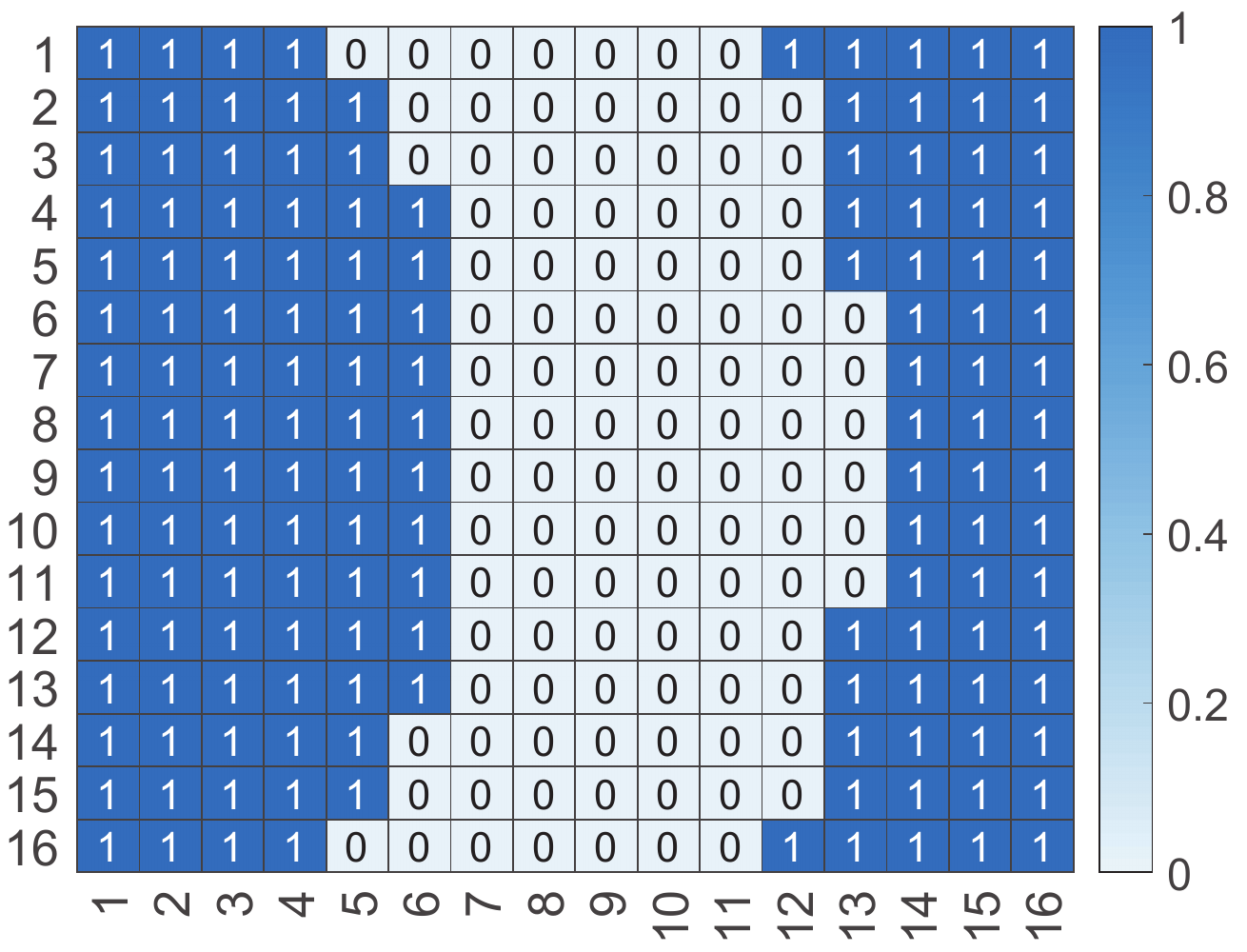}} \\

    \subfigure[]{
		\includegraphics[width=0.35\linewidth]{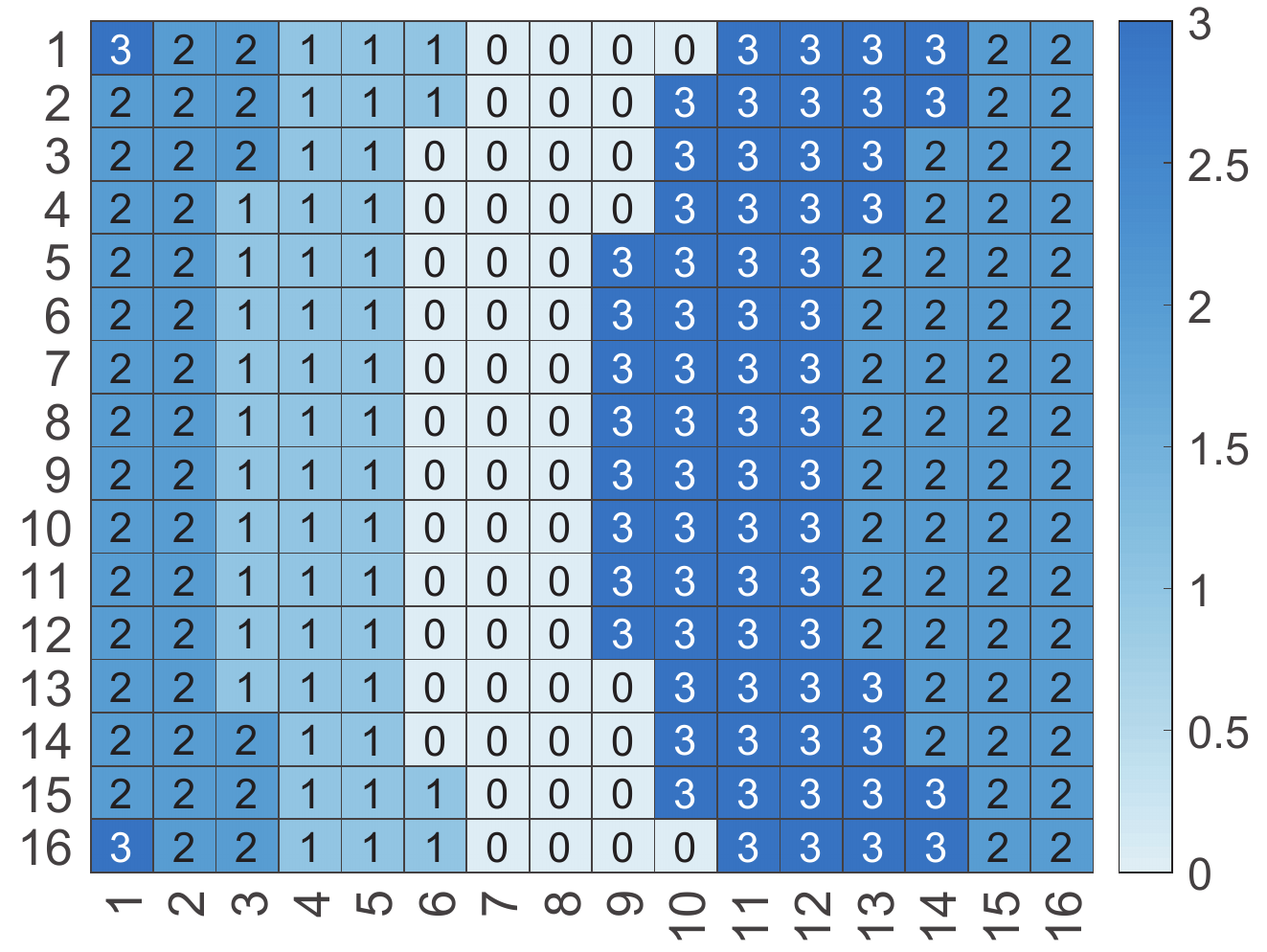}}
    \subfigure[]{
		\includegraphics[width=0.35\linewidth]{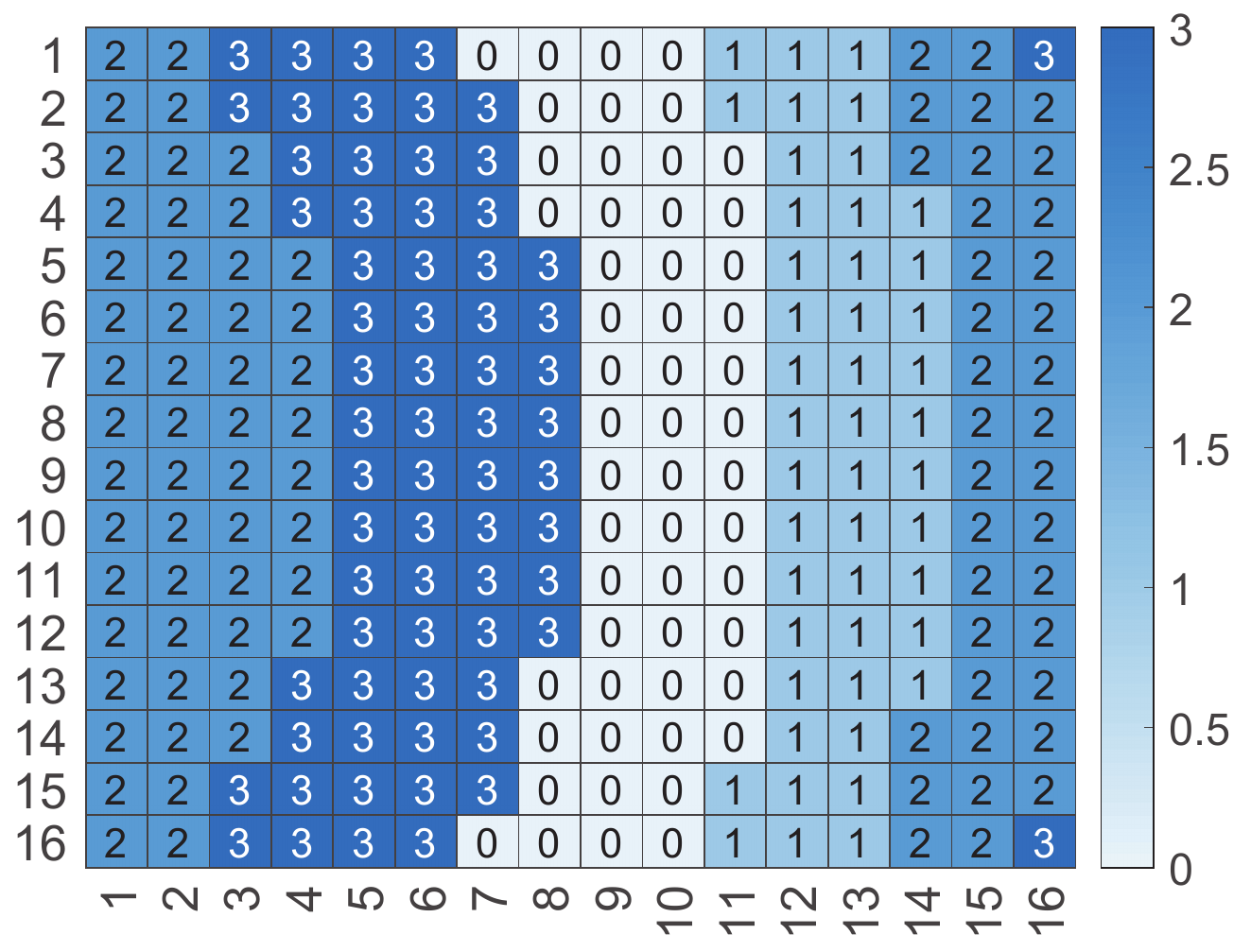}}
	\caption{The RIS codeword. (a) $(\theta_r,\phi_r)=(10^{\circ}, 0^{\circ})$ under one-bit quantization. (b) $(\theta_r,\phi_r)=(10^{\circ}, 180^{\circ})$ under one-bit quantization. (c) $(\theta_r,\phi_r)=(10^{\circ}, 0^{\circ})$ under two-bit quantization. (d) $(\theta_r,\phi_r)=(10^{\circ}, 180^{\circ})$ under two-bit quantization.}
\label{codeword}
\end{figure}
The measured results are displayed in Fig. \ref{imaging_result}.
For ease of notation, we utilize the minus of $\theta$ within the figure to represent that the receiver is located at the mirror image of $(\theta, 0^{\circ})$, i.e., $(\theta, 180^{\circ})$.
On the one hand, we can see that when $d_t$ is set as 3.6m, the incident signal is also focused towards $(10^{\circ},180^{\circ})$., i.e., the imaging interference appears.
Conversely, when $d_t$ is set as 0.5m, the imaging interference almost disappears, which demonstrates that $d_t$ has a significant impact on the imaging interference.
On the other hand, when $d_t$ is set as 3.6m, it can be checked that the received power at $(10^{\circ},180^{\circ})$ is almost the same as that at $(10^{\circ}, 0^{\circ})$ for one-bit quantization.
However, the received power at $(10^{\circ},180^{\circ})$ is about 10 dBm lower than that at $(10^{\circ}, 0^{\circ})$ for two-bit quantization, which demonstrates that the quantization bit number also has an impact on the imaging interference.
To interpret this phenomenon in an intuitive manner, we introduce a new concept of {\it codeword similarity} $\mu_b[(\theta_1, \phi_1),(\theta_2, \phi_2)]$ to represent the proportion of the RIS units with the same phase shift between two codewords, where one is designed to reflect the incident signal towards $(\theta_1, \phi_1)$ and the other is designed to reflect the incident signal towards $(\theta_2, \phi_2)$.
The subscript $b$ denotes the quantization bit number, which is equal to 1 or 2.
The RIS codewords under both one-bit quantization and two-bit quantization are shown in Fig. \ref{codeword}.
We can see that $\mu_2[(10^{\circ}, 0^{\circ}),(10^{\circ}, 180^{\circ})]=0.32$ is quite smaller than $\mu_1[(10^{\circ}, 0^{\circ}),(10^{\circ}, 180^{\circ})]=0.75$, which explains why two-bit phase quantization can alleviate the imaging interference.
\subsection{Multi-beam Reflection}
In this subsection, the following case is considered in order to demonstrate the RIS's multi-beam reflection performance.

{\bf Case VIII:}
The RIS is designed to reflect the incident signal towards the two different directions simultaneously.
Then, we test the angle power spectrum of the reflected signal.

\begin{figure}[!t]
	\centering
	\subfigbottomskip=2pt
	\subfigcapskip=-5pt
	\subfigure[]{
		\includegraphics[width=0.41\linewidth]{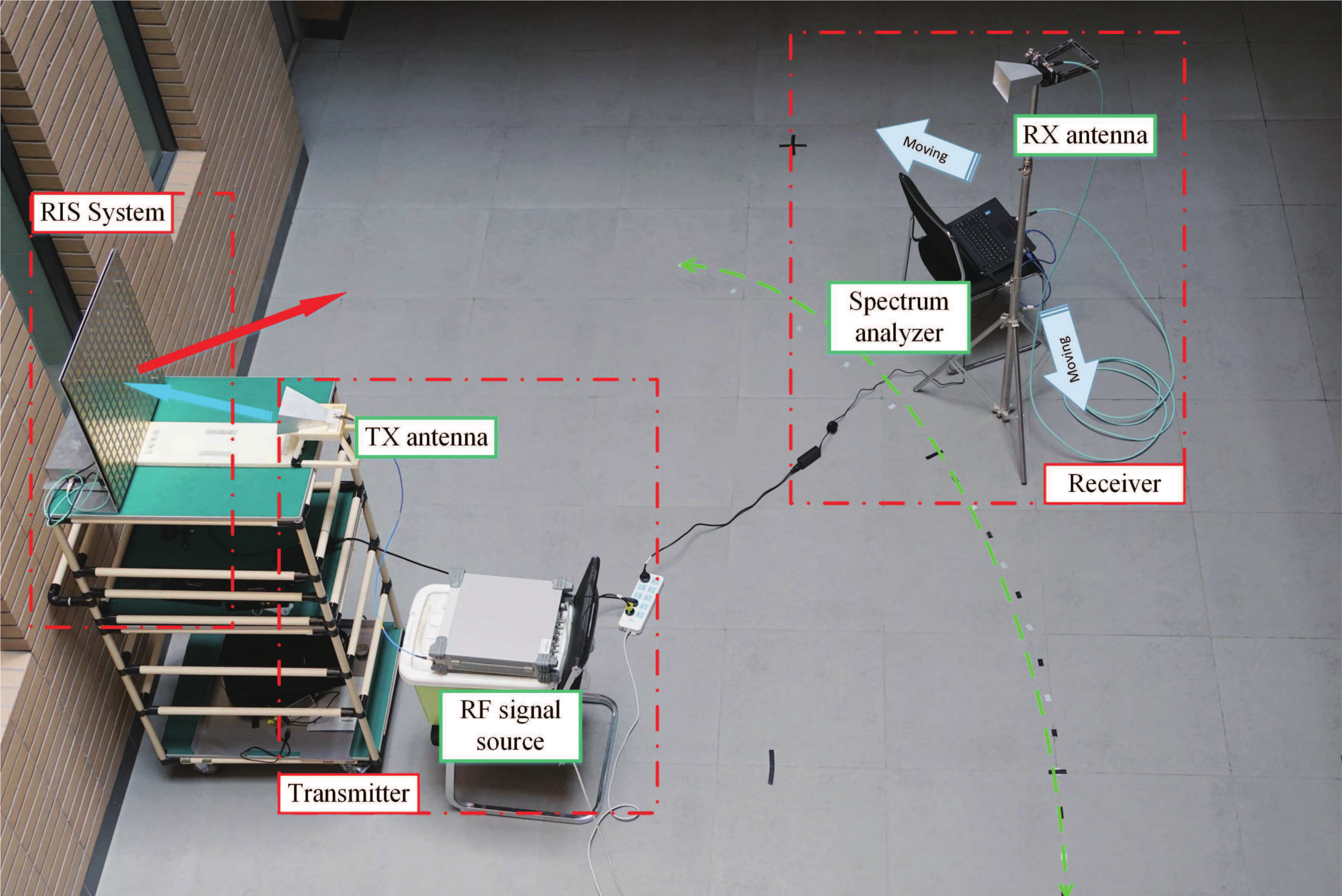}}
	\subfigure[]{
		\includegraphics[width=0.56\linewidth]{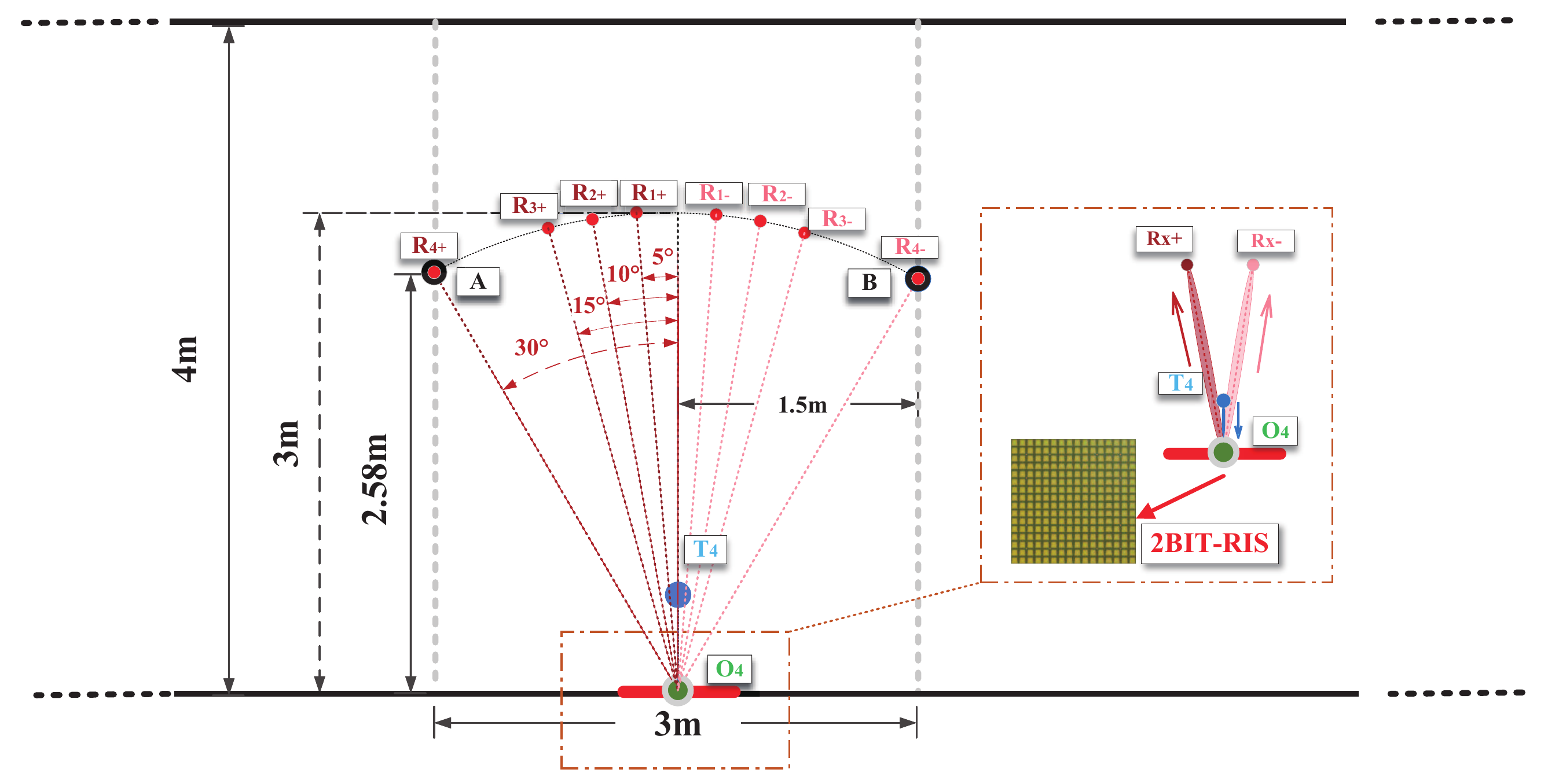}}
\caption{Multi-beam reflection measurement. (a) Test environment. (b) Plane topology.}
\label{multibeam_environment}
\end{figure}

Since $d_t$ is set as 0.5m, there is no imaging interference in this case, which has been proved in Section IV-F.
The test environment and the plane topology are separately shown in Fig. \ref{multibeam_environment}(a) and Fig. \ref{multibeam_environment}(b).
The RIS is located at point `$\text O_4$' and the transmitter is located at point `$\text T_4$'.
We test the angle power spectrum of the received signal by moving the receiver along the green line in Fig. \ref{multibeam_environment}(a), which corresponds to the arc $\overset{\frown} {\text{AB}}$ in Fig. \ref{multibeam_environment}(b).
Taking the RIS's dual-beam as an example, we consider four sets of target angles, including $\{(5^{\circ},0^{\circ}), (5^{\circ},180^{\circ})\}$, $\{(10^{\circ},0^{\circ}), (10^{\circ},180^{\circ})\}$, $\{(15^{\circ},0^{\circ}), (15^{\circ},180^{\circ})\}$ and $\{(30^{\circ},0^{\circ}), (30^{\circ},180^{\circ})\}$.
Note that these four sets of target angles are separately correspond to the sets of test points $\{\text R_{1+}, \text R_{1-}\}$, $\{\text R_{2+}, \text R_{2-}\}$, $\{\text R_{3+}, \text R_{3-}\}$ and $\{\text R_{4+}, \text R_{4-}\}$ in Fig. \ref{multibeam_environment}(b).
Besides, the PA technology \eqref{multibeam} is utilized to generate the RIS codebooks.
\begin{figure}[!t]
	\centering	\includegraphics[width=105mm]{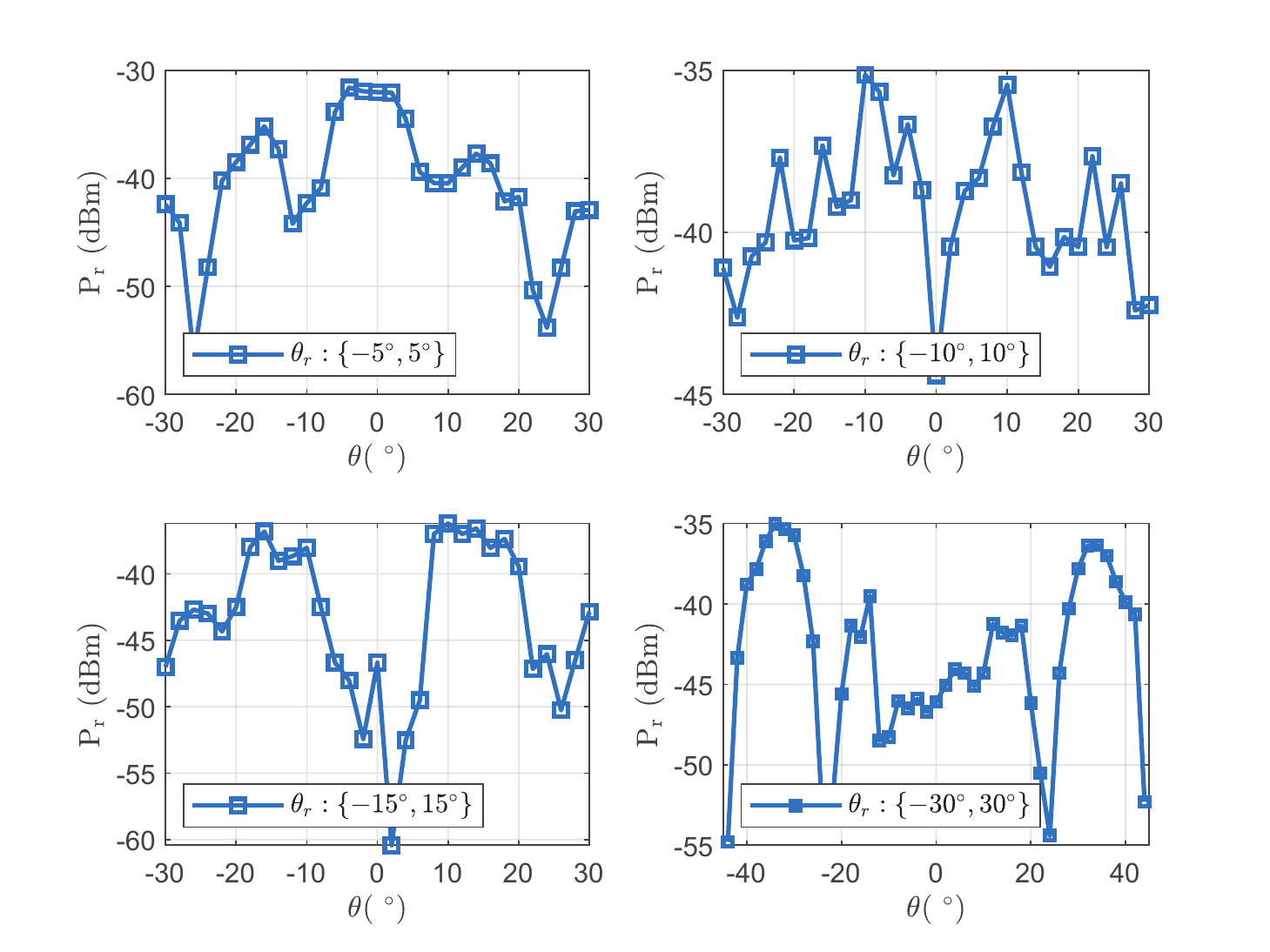}
	\caption{The angular power spectrum in terms of RIS multi-beam reflection.}
	\label{Fig:fig_Multibeam}
\end{figure}
The theoretical beamwidth $B_w$ of the RIS can be computed as \cite{190}
\begin{align}
         B_w = 2\left[\frac{\pi}{2}-\cos^{-1}\left(\frac{1.391\lambda}{\pi N d}\right)\right]=7.695^{\circ},
\end{align}
where $N=16$ and $d = 0.408\lambda$.
Ideally, $B_w$ can be utilized to approximate the beam resolution of the RIS.
Fig. \ref{Fig:fig_Multibeam} shows the angle power spectrum of the reflected signal, where the minus of $\theta$ means that the receiver is located at the mirror image of $(\theta, 0^{\circ})$, i.e., $(\theta, 180^{\circ})$.
When the target beams  are $\{(5^{\circ},0^{\circ}), (5^{\circ},180^{\circ})\}$, two beams is merged and can not be distinguished from each other.
However, two clearly distinguishable beams can be observed when target angles are $\{(10^{\circ},0^{\circ}), (10^{\circ},180^{\circ})\}$, $\{(15^{\circ},0^{\circ}), (15^{\circ},180^{\circ})\}$ and $\{(30^{\circ},0^{\circ}), (30^{\circ},180^{\circ})\}$.
\subsection{Mutual Coupling}
In this subsection, the following case is considered in order to demonstrate the existence and the effect of the mutual coupling among different RIS units.

{\bf Case IX:}
The RIS is designed to reflect the incident signal towards the receiver.
Then, we compare the practical received power to its theoretical counterpart without the consideration of the mutual coupling.
Note that we separately place the RIS against the wall and not against the wall, which aims at eliminating the effects of the surrounding environment.

Taking account of the mutual coupling between different RIS units, the authors in \cite{19} have proposed an end-to-end channel model of the RIS-aided system, which is written as
\begin{align}\label{coupling}
        h=\beta_0\mathbf h^T_{rx}(\mathbf Z_{RIS}+\mathbf Z_{ss})^{-1}\mathbf h_{tx},
\end{align}
where $\beta_0$ is a constant, $\mathbf h_{tx}$ and $\mathbf h_{rx}$ have been introduced in Section II-A, $\mathbf Z_{RIS}$ denotes the adjustable diagonal impendence matrix of the RIS that is similar to $\mathbf \Theta$, and $\mathbf Z_{ss}$ denotes the mutual coupling matrix of the RIS, which can be determined by the Green's function.
We can see that if the mutual coupling does not exist, i.e., each element of $\mathbf Z_{ss}$ is zero, then the expression of \eqref{coupling} will degenerate to that in \eqref{channel}.

However, there is no existing work that has validated the existence of the mutual coupling between different RIS units.
To accomplish this, we propose a strategy as follows.
We denote the proportion of the ``invalid" RIS units as $r_i$. 
Specifically, if the RIS unit is ``valid", then it is designed according to \eqref{singlebeam} to reflect the incident signal towards the receiver.
On the contrary, if the RIS unit is ``invalid", then its phase shift is set as a default value, e.g., $90^{\circ}$.
Note that all the ``valid" units contribute to the desired exit beam, while all the ``invalid" units act as a metal plate, which perform specular reflection and result in the specular reflection beam.
If the mutual coupling does not exist, then the received power $P_r$ is due to the contributions of all the RIS units and its theoretical value can be computed by \eqref{Pr}.
Thus, we set different values of $r_i$ and compute the theoretical value of $P_r$ versus $r_i$.
Next, we measure the practical value of $P_r$ versus $r_i$.
Note that if the mutual coupling does not exist, the measured $P_r$ versus $r_i$ should broadly match the theoretical counterpart in terms of the curve variation trend.
In other words, if the variation trends of the theoretical and the measured values of $P_r$ differ relative greatly, then it means that the received power is not completely determined by the sum of the contribution of all the RIS units, which indirectly proves the existence of the mutual coupling effect.

\begin{figure}[!t]
	\centering
	\subfigbottomskip=2pt
	\subfigcapskip=-5pt
	\subfigure[]{
		\includegraphics[width=0.41\linewidth]{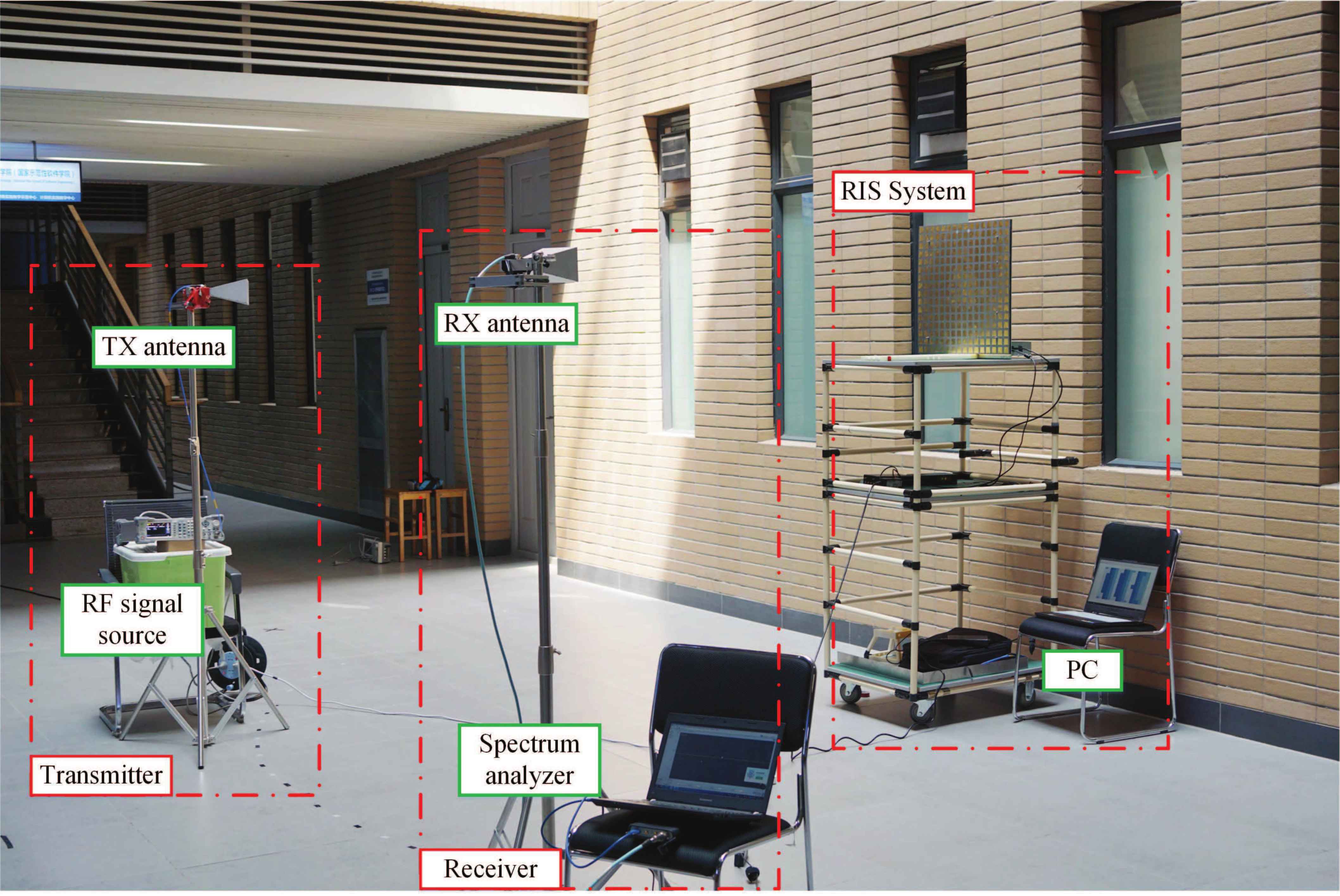}}
	\subfigure[]{
		\includegraphics[width=0.56\linewidth]{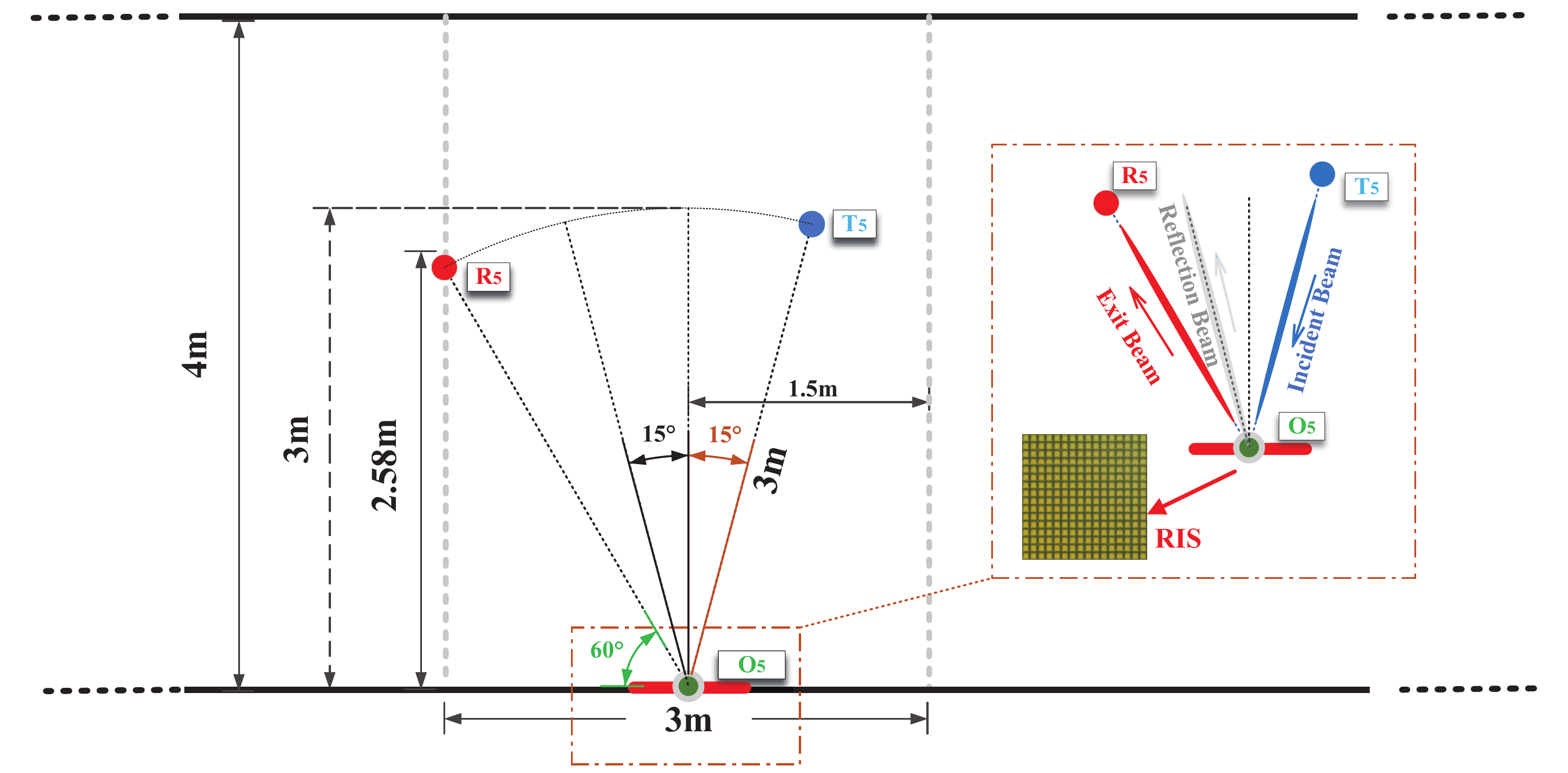}}
\caption{Mutual coupling measurement (the RIS is against the wall). (a) Test environment. (b) Plane topology.}
\label{diagram111}
\end{figure}
\begin{figure}[!t]
	\centering
	\subfigbottomskip=2pt
	\subfigcapskip=-5pt
	\subfigure[]{
		\includegraphics[width=0.41\linewidth]{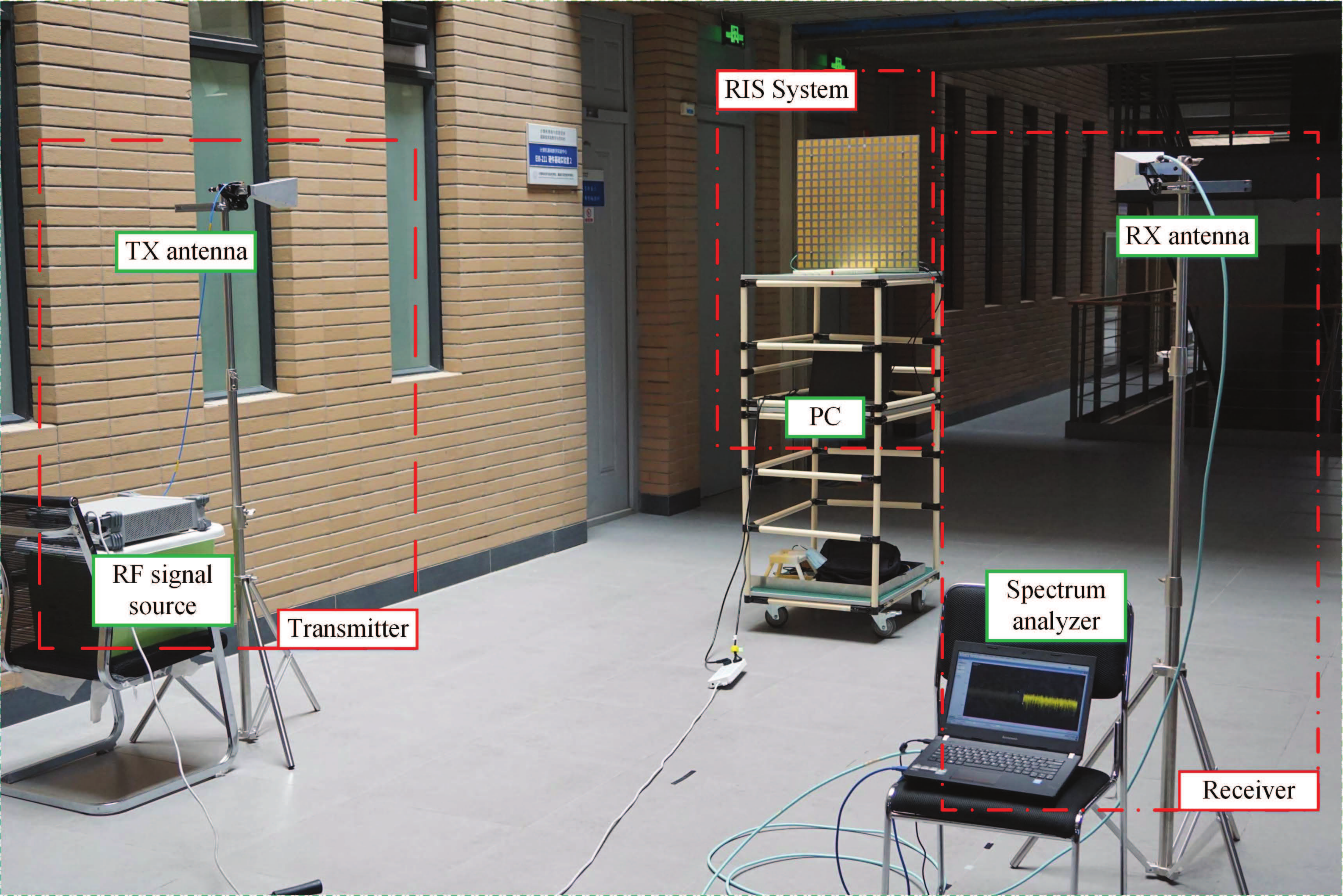}}
	\subfigure[]{
		\includegraphics[width=0.56\linewidth]{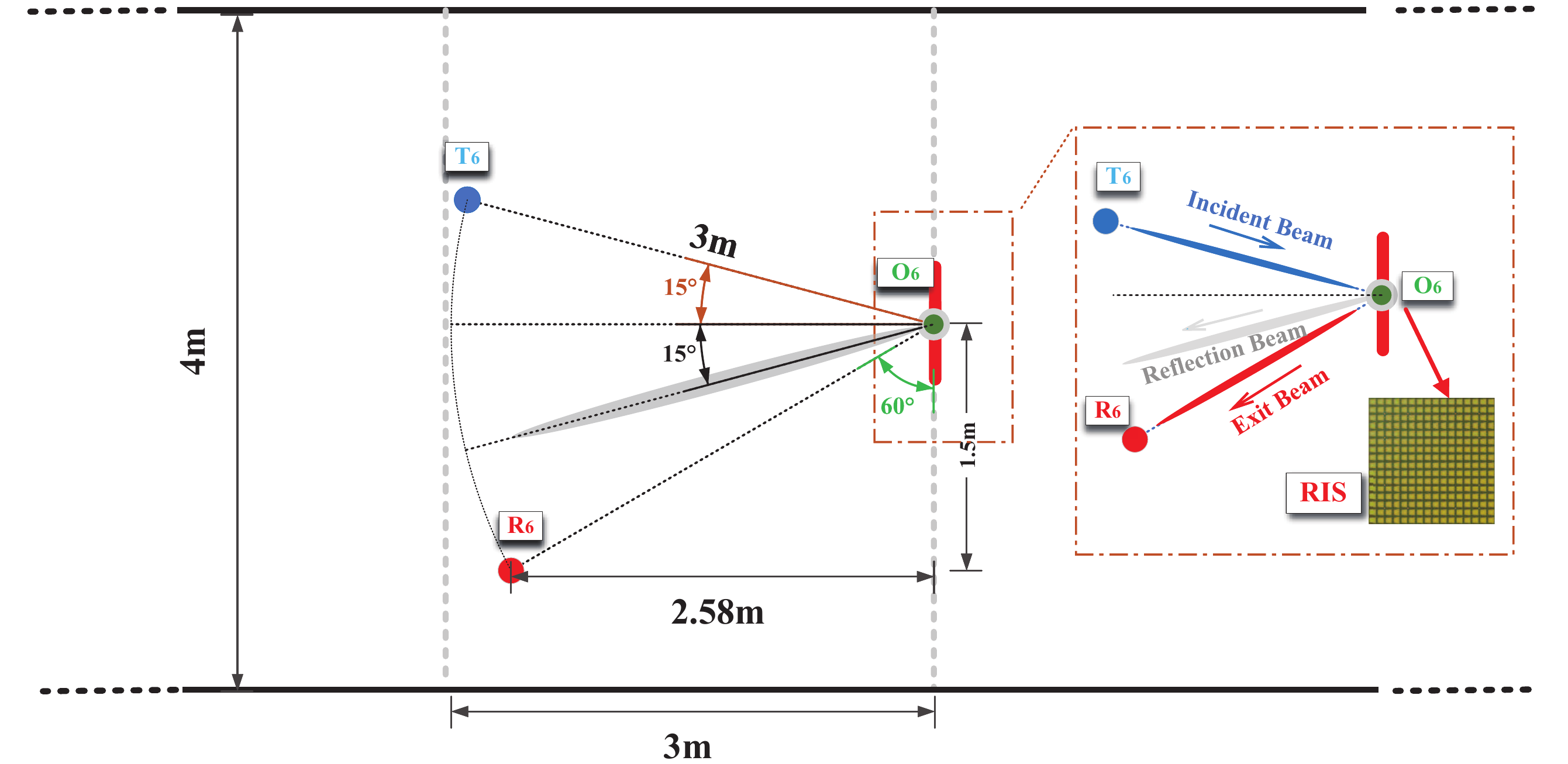}}
\caption{Mutual coupling measurement (the RIS is not against the wall). (a) Test environment. (b) Plane topology.}
\label{diagram1111}
\end{figure}
It must be pointed out that the interference induced by the surrounding environment should be considered, which mainly involves the surrounding wall.
To evaluate the effects of the surrounding wall, we test the measured received power under two scenarios, where one corresponds to the scene that the RIS is placed against the wall and the other corresponds to the scene that their is no wall near the RIS.
The test environment and the plane topology of the former case are separately shown in Fig. \ref{diagram111}.(a) and Fig. \ref{diagram111}.(b), while those of the latter case are separately illustrated in Fig. \ref{diagram1111}.(a) and Fig. \ref{diagram1111}.(b).
The RIS is positioned at point `$\text O_5$' and `$\text O_6$', the transmitter is positioned at point `$\text T_5$' and point `$\text T_6$', while the receiver is positioned at point `$\text R_5$' and point `$\text R_6$'.
Moreover, we set $d_t=3$m, $h_t=1.8$m, $\theta_i=15^{\circ}$ and $\phi_i=180^{\circ}$ in this case.

\begin{figure}[!t]
	\centering
	\includegraphics[width=120mm]{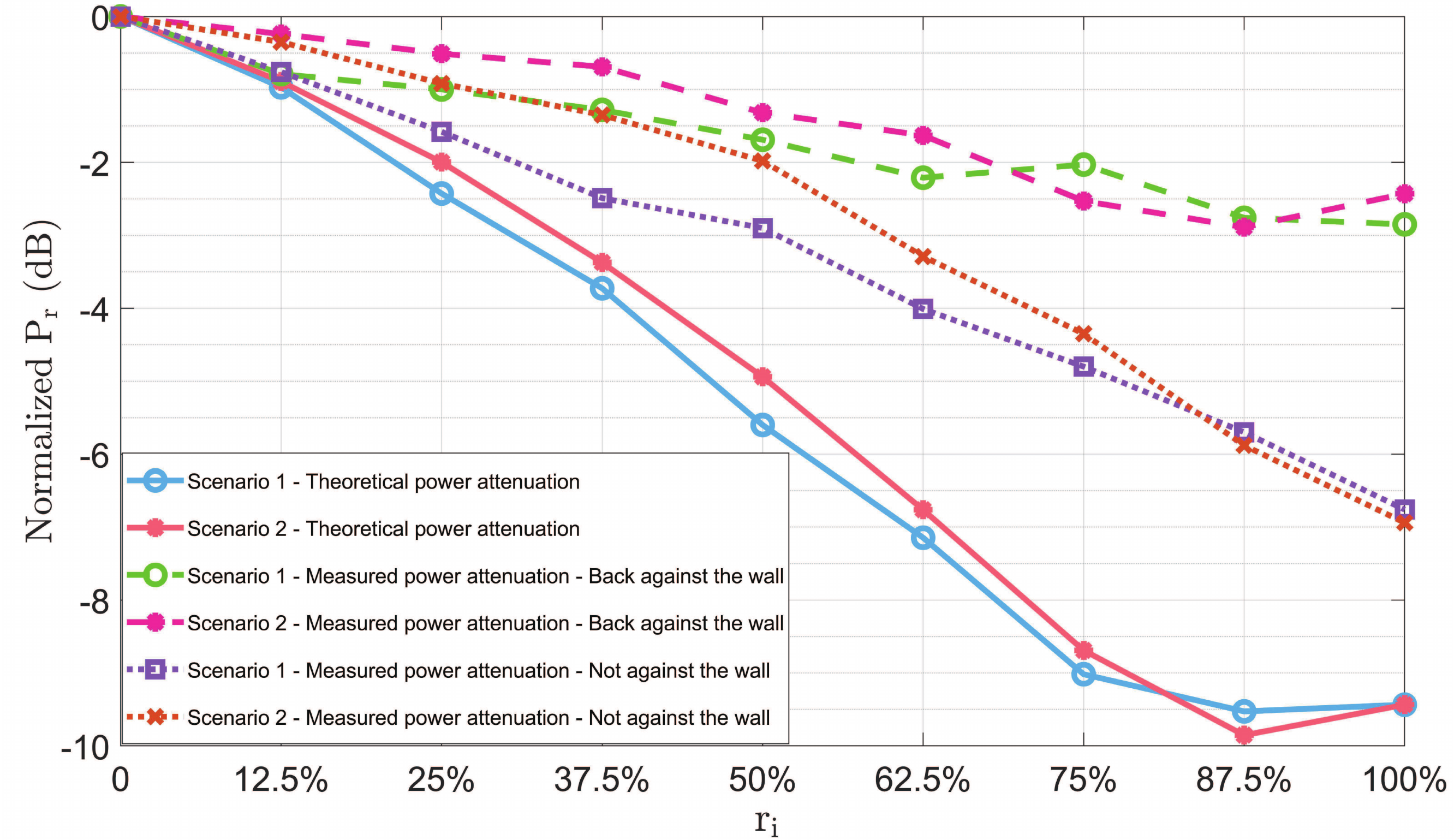}
	\caption{{The received power $P_r$ versus $r_i$.}}
	\label{coupling_result}
\end{figure}
The normalized received power $P_r$ versus $r_i$ is shown in Fig. \ref{coupling_result}.
The scenario 1 and scenario 2 correspond to different location distribution of the ``invalid'' units.
We can find that the effect of the surrounding environment is less than 3 dB.
Then, we focus on the case where the RIS is placed not against the wall.
It can be observed that there is a gap between the theoretical $P_r$ and the measured one.
In other words, the practical received power is not exactly equal to the sum of the contribution of all RIS units.
Specifically, if $r_i=87.5\%$, then the gap is about 4 dB.
Since the effect of the surrounding environment has been excluded, this gap can only be because there exists mutual coupling between different RIS units.
Specifically, a total of 256 RIS units result in a maximum of 4 dB power gain due to the mutual coupling.
Moreover, as $r_i$ increases, the gap increases correspondingly.
This is because that as the proportion of the ``invalid'' units increases, the contribution of RIS units will weaken, and the effect of the mutual coupling will become dominant gradually.
Furthermore, we change the location of the ``valid" units with their number fixed, and it can be checked that its has little influence to $P_r$.

\section{Conclusion}

In this paper, we have proposed a two-bit RIS-aided communication system, where a two-bit RIS is deployed to aid the communication from a single-antenna transmitter to a single-antenna receiver.
Moreover, a corresponding prototyping system has been designed to accomplish experimental tests in practical environments.
A fully self-developed communication terminal has been adopted, which can fully acquire the channel information.
We have tested and validated a few functions and properties of the RIS as follows.
Firstly, we have validated a novel power consumption model of the RIS, which takes account of both the static and the dynamic power consumption.
Secondly, we have demonstrated the existence of the imaging interference in RIS-aided systems, and we have found that two-bit RIS can effectively suppress the imaging interference.
Moreover, we have tested that the RIS can outperform the metal plate in terms of both the beam focusing capacity and the channel stationarity improvement.
Lastly, we have validated the multi-beam reflection of the RIS and verified the existence of the mutual coupling between different RIS units.

\linespread{1.5}


\begin{thebibliography}{1}
\bibitem{1}
T. J. Cui, M. Q. Qi, X. Wan, J. Zhao and Q. Cheng, ``Coding metamaterials, digital metamaterials, and programmable metamaterials," \emph{Light Sci. Appl.,} vol. 3, pp. 1--9, Oct. 2014.
\bibitem{2}
M. Di Renzo \emph{et al.}, ``Smart radio environments empowered by reconfigurable intelligent surfaces: How it works, state of research, and the road ahead," \emph{IEEE J. Sel. Areas Commun.,} vol. 38, no. 11, pp. 2450--2525, Nov. 2020.
\bibitem{3}
S. Zhang, S. Zhang, F. Gao, J. Ma and O. A. Dobre, ``Deep learning optimized sparse antenna activation for reconfigurable intelligent surface assisted communication,"  \emph{IEEE Trans Commmu.,} vol. 69, no. 10, pp. 6691-6705, Oct. 2021.
\bibitem{4}
W. Tang, M. Z. Chen, J. Y. Dai, Y. Zeng, X. Zhao, S. Jin, Q. Cheng and T. J. Cui, ``Wireless communications with programmable metasurface: New paradigms, opportunities, and challenges on transceiver design,"  \emph{IEEE Wireless Commun.,} vol. 27, no. 2, pp. 180--187, Apr. 2020.
\bibitem{30}
S. Zhang, M. Li, M. Jian, Y. Zhao and F. Gao, ``AIRIS: Artificial intelligence enhanced signal processing in reconfigurable intelligent surface communications,"  arXiv:2106.00171, 2021. [Online].~Available: https://arxiv.org/abs/2106.00171.

\bibitem{31}
C. Huang, A, Zappone, G. C. Alexandropoulos, M. Debbah and C. Yuen ``Reconfigurable intelligent surfaces for energy efficiency in wireless communication,"  \emph{IEEE Trans. Wireless  Commmu.,} vol. 18, no. 8, pp. 4157--4170, Aug. 2019.
\bibitem{32}
Y. Han, W. Tang, S. Jin, C. K. Wen and X. Ma, ``Large intelligent surface-assisted wireless communication exploiting
statistical CSI," \emph{IEEE Trans. Veh. Technol.,} vol. 68, no. 8, pp. 8238--8242, Aug. 2019.

\bibitem{100}
X. Wei, L. Dai, Y. Zhao, G. Yu and X. Duan, ``Codebook design and beam training for extremely large-scale RIS: Far-field
or near-field?" \emph{China Commun.,} Jun. 2021.


\bibitem{7}
J. Wang \emph{et al.}, ``Reconfigurable intelligent surface: Power consumption modeling and practical measurement validation," arXiv:2211.00323, 2022. [Online].~Available: https://arxiv.org/abs/2211.00323.
\bibitem{8}
J. Wang, W. Tang, S. Jin, X. Li and M. Matthaiou, ``Static power consumption modeling and measurement of reconfigurable intelligent surfaces," arXiv:2303.00299, 2023. [Online].~Available: https://arxiv.org/abs/2303.00299.

\bibitem{9}
G. C. Alexandropoulos, V. Jamali, R. Schober and H. V. Poor, ``Near-field hierarchical beam management for RIS-enabled millimeter wave multi-antenna systems," arXiv:2203.15557, 2022. [Online].~Available: https://arxiv.org/abs/2203.15557.
\bibitem{10}
M. Ouyang, Y. Wang, F. Gao, S. Zhang, P. Li and J. Ren, ``Computer vision-aided reconfigurable intelligent surface-based beam tracking: Prototyping and experimental results," arXiv:2207.05032, 2022. [Online].~Available: https://arxiv.org/abs/2207.05032.
\bibitem{200}
M. Mizmizi, D. Tagliaferri and U. Spagnolini, ``Wireless communications with space-time modulated metasurfaces," arXiv:2302.08310, 2023. [Online].~Available: https://arxiv.org/abs/2302.08310.


\bibitem{14}
N. M. Tran, M. M. Amri, J. H. Park, D. I. Kim and K. W. Choi, ``Multifocus techniques for reconfigurable intelligent surface-aided wireless power transfer: Theory to experiment," \emph{IEEE Internet Things J.,} vol. 9, no. 18, pp. 17157--17171, Sep. 2022.
\bibitem{15}
Y. Shuang, H. Zhao, M. Wei, Q. Cheng, S. Jin, T.J. Cui, P. d. Hougne and L. Li, ``One-bit quantization is good for programmable coding metasurfaces," \emph{Science China Information Sciences,} 65(7):1--15, 2022.

\bibitem{5}
J. Hu, H. Yin, L. Tan, L. Cao and X. Pei, ``RIS-aided wireless communications: Can RIS beat metal plate?" arXiv:2303.02938, 2023. [Online].~Available: https://arxiv.org/abs/2303.02938.

\bibitem{6}
O. Ozdogan, E. Bjornson, and E. G. Larsson, ``Intelligent reflecting surfaces: Physics, propagation, and pathloss modeling," \emph{IEEE Wireless Comm. Let.,} vol. 9, no. 5, pp. 581--585, May. 2020.
\bibitem{11}
W. Tang \emph{et al.}, ``Wireless communications with reconfigurable intelligent surface: Path loss modeling and experimental measurement," \emph{IEEE Trans. Wireless Comm.,} vol. 20, no. 1, pp. 421--439, Jan. 2021.
\bibitem{12}
W. Tang \emph{et al.}, ``Path loss modeling and measurements for reconfigurable intelligent surfaces in the millimeter-wave frequency band," \emph{IEEE Trans. Commun.,} vol. 70, no. 9, pp. 6259--6276, Sep. 2022.
\bibitem{13}
Z. Wang, L. Tan, H. Yin, K. Wang, X. Pei and D. Gesbert ``A received power model for reconfigurable intelligent surface and measurement-based validations," arXiv:2211.00323, 2022. [Online].~Available: https://arxiv.org/abs/2211.00323.
\bibitem{21}
M. Najafi, V. Jamali, R. Schober, and H. V. Poor, ``Physics-based modeling and scalable optimization of large intelligent reflecting surfaces," \emph{IEEE Trans. Wireless Comm.,} vol. 69, no. 4, pp. 2673-2691, Apr. 2021.
\bibitem{22}
F. H. Danufane \emph{et al.}, ``On the path-loss of reconfigurable intelligent surfaces: An approach based on Green's theorem applied to vector fields," \emph{IEEE Trans. Wireless Comm.,} vol. 69, no. 8, pp. 5573-5592, Aug. 2021.

\bibitem{23}
D. Radovic, F. Pasic, M. Hofer, H. Groll, C. F. Mecklenbraucker and T. Zemen, ``Stationarity evaluation of high-mobility sub-6 GHz and mmWave non-WSSUS channels," arXiv:2304.00870, 2023. [Online].~Available: https://arxiv.org/abs/2304.00870.

\bibitem{26}
H. Sun , S. Zhang, J. Ma and O. A. Dobre, ``Time-delay unit based beam squint mitigation for RIS-aided communications," \emph{IEEE Commun. Lett.,} vol. 26, no. 9, pp. 2220--2224, Sept. 2022.

\bibitem{16}
H. Taghvaee, A. Jain, S. Abadal, G. Gradoni, E. Alarcon and A. C. Aparicio, ``On the enabling of multi-receiver communications with reconfigurable intelligent surfaces,"  \emph{IEEE Transactions on Nanotechnology,} vol. 21, pp. 413--423, Jul. 2022.
\bibitem{17}
L. Bao, R. Y. Wu, X. Fu, Q. Ma, G. D. Bai, J. Mu, R. Jiang and T. J. Cui, ``Multi-beam forming and controls by metasurface with phase and amplitude modulations," \emph{IEEE Trans. Antennas Propag.,} vol. 67, no. 10, pp. 6680--6685, Jul. 2019.
\bibitem{18}
N. Torkzaban \emph{et al.}, ``Shaping mmWave wireless channel via multi-beam design using reconfigurable intelligent surfaces," arXiv:2110.00101, 2021. [Online].~Available: https://arxiv.org/abs/2110.00101.

\bibitem{38}
F. Xiao, W. Liu and Y. Kami, ``Analysis of crosstalk between finite-length microstrip lines: FDTD approach and circuit-concept modeling," \emph{IEEE Trans. Electromagn. Compat.,} vol. 43, no. 4, pp. 573--578, Nov. 2001.
\bibitem{39}
L. Yan, X. Zhang, X. Zhao, X. Zhou and R. X. Gao, ``A fast and efficient analytical modeling approach for external electromagnetic field coupling to transmission lines in a metallic enclosure," \emph{IEEE Access,} vol. 6, pp. 50272--50277, 2018.

\bibitem{19}
G. Gradoni and M. D. Renzo, ``End-to-end mutual coupling aware communication model for reconfigurable intelligent surfaces: An electromagnetic-compliant approach based on mutual impedances," \emph{IEEE Wireless Commun. Lett.,}  vol. 10, no. 5, pp. 938--942, May. 2019.

\bibitem{190}
W. L. Stutzman and G. A. Thiele, {\it Antenna Theory and Design}, 3rd ed., New York, USA: Wiley, 2012.
\end{thebibliography}
\end{document}